\documentclass[aps,prd,reprint,amsmath,amssymb,showpacs,nofootinbib,onecolumn,notitlepage]{revtex4-1}
\usepackage{epsfig,slashed,xcolor,bm,natbib,comment,enumitem,ulem,cancel}
\usepackage{calrsfs}
\DeclareMathAlphabet{\pazocal}{OMS}{zplm}{m}{n}
\usepackage[cal=boondox]{mathalfa}

\newcommand{\Db}{\pazocal{D}}
\newcommand{\Tb}{\pazocal{T}}

\def\pd#1#2{\frac{\partial #1}{\partial #2}}
\def\bpd#1#2{\frac{\bar{\partial} #1}{\partial #2}}

\def\tfrac#1#2{{\textstyle\frac{#1}{#2}}}

\def\momc#1#2{\pd{{\cal L}}{(\partial_{#1}\chi_{#2})}}
\def\momf#1#2{\pd{{\cal L}}{(\partial_{#1} #2)}}

\newcommand{\vpsi}{\varphi}

\def\zero#1{{^0}\!#1}
\def\czero#1{{^0}#1}

\newcommand{\comma}{, }
\newcommand{\be}{\begin{equation}}
\newcommand{\ee}{\end{equation}}
\newcommand{\bea}{\begin{eqnarray}}
\newcommand{\eea}{\end{eqnarray}}
\newcommand{\myDel}[1]{{\color{red}\ifmmode\cancel{#1}\else\sout{#1}\fi}}

\newcommand{\eom}{\bumpeq}
\newcommand{\eomm}{\stackrel{\mbox{\tiny m}}{\bumpeq}}
\newcommand{\eomg}{\stackrel{\mbox{\tiny g}}{\bumpeq}}
\begin{document}
\title{Manifestly covariant variational principle for gauge theories
  of gravity}
\author{M.P.~\surname{Hobson}}
\email{mph@mrao.cam.ac.uk} 
\author{A.N.~\surname{Lasenby}}
\email{anthony@mrao.cam.ac.uk}
\author{W.E.~\surname{Barker}}
\email{wb263@cam.ac.uk}
\affiliation{Astrophysics Group, Cavendish
  Laboratory, JJ Thomson Avenue, Cambridge CB3 0HE\comma UK}
\begin{abstract}
A variational principle for gauge theories of gravity is presented,
which maintains manifest covariance under the symmetries to which the
action is invariant, throughout the calculation of the equations of
motion and conservation laws. This is performed by deriving explicit
manifestly covariant expressions for the Euler--Lagrange variational
derivatives and Noether's theorems for a generic action of the form
typically assumed in gauge theories of gravity. The approach is
illustrated by application to two scale-invariant gravitational gauge
theories, namely Weyl gauge theory (WGT) and the recently proposed
`extended' Weyl gauge theory (eWGT), where the latter may be
considered as a novel gauging of the conformal group, but the method can
be straightforwardly applied to other theories with smaller or larger
symmetry groups. The approach also enables one easily to establish the
relationship between manifestly covariant forms of variational
derivatives obtained when one or more of the gauge field strengths is
set to zero either before or after the variation is performed. This is
illustrated explicitly for both WGT and eWGT in the case where the
translational gauge field strength (or torsion) is set to zero before
and after performing the variation, respectively.
\end{abstract}


\maketitle

\section{Introduction}
\label{sec:intro}

For any given action, the process of deriving the manifestly covariant
equations of motion for the fields on which it depends can be very
time-consuming. A key reason is that for an action that is invariant
under some set of symmetries, either global or local, the individual
terms making up the Euler--Lagrange equations are typically not
covariant under those symmetries. One therefore usually obtains
equations of motion that, although inevitably covariant, are not
manifestly so. One then faces the task of combining terms in various
ways to achieve manifest covariance before continuing with further
analysis, and this process can require considerable trial and error,
often relying on inspired guesswork. Similar difficulties are also
encountered when deriving conservation laws, which must again be
covariant under the symmetries of the action, but are typically not
obtained in a manifestly covariant form when they are derived using
the standard forms of Noether's theorems.

Here we present an alternative approach whereby one maintains manifest
covariance throughout the calculation of the equations of motion and
conservation laws, thereby circumventing the above
difficulties. Methods for achieving this, at least for the equations
of motion, have been considered previously in the context of
gravitational theories that are interpreted in the usual geometrical
manner, where the action depends typically on the spacetime metric
$g_{\mu\nu}$, together perhaps with some non-metric connection
${\Gamma^\sigma}_{\mu\nu}$
\cite{Exirifard08,Borunda08,Dadich11,Cremaschini15,Cremaschini16,Tessarotto21}.
Here we instead focus on developing a manifestly covariant variational
principle for gauge theories of gravity
\cite{Utiyama56,Kibble61,Sciama64,Hehl76,Blagojevic02}.  In
particular, we illustrate the method by application to the
scale-invariant Weyl gauge theory (WGT)
\cite{Dirac73,Bregman73,Charap74,Kasuya75,Omote77,Sijacki82,Neeman88}
and its recently proposed `extended' version (eWGT)
\cite{eWGTpaper,CGTpaper}, but the approach presented can be
straightforwardly applied to other theories with smaller or larger
symmetry groups, such as Poincar\'e gauge theory (PGT)
\cite{Hehl76,vonderheyde76,Hayashi80,Obukhov89} or conformal gauge
theory (CGT)
\cite{Crispim77,Crispim78,Kaku77,Kaku78,Lord85,Wheeler91}. In
addressing WGT and eWGT, we assume the action to depend on a
translational gauge field ${h_a}^\mu$, a rotational gauge field
${A^{ab}}_\mu$ and a dilational gauge field $B_\mu$, together with
some set of matter fields $\vpsi_A$, which may include a scalar
compensator field (which we occasionally also denote by $\phi$).  It
is worth noting that gauge theories of gravitation are most naturally
interpreted as field theories in Minkowski spacetime
\cite{Wiesendanger96,Lasenby98}, in the same way as the gauge field
theories describing the other fundamental interactions, and this is
the viewpoint that we shall adopt here. It is common, however, to
reinterpret the mathematical structure of gravitational gauge theories
geometrically, where in particular the translational gauge field
${h_a}^\mu$ is considered as forming the components of a vierbein (or tetrad)
system in a more general Weyl--Cartan spacetime, in which
${A^{ab}}_\mu$ and $B_\mu$ then correspond to the spin-connection and
Weyl vector, respectively \cite{Hehl76}.  These issues are discussed
in more detail elsewhere \cite{Blagojevic02,eWGTpaper}.

The manifestly covariant approach presented here also enables one
easily to establish the relationship between the forms of variational
derivatives, and hence the field equations, obtained by applying
first- and second-order variational principles, respectively. A
particularly interesting case is provided by comparing the variational
derivatives obtained by setting the translational gauge field strength
(or torsion) to zero after the variation is performed (first-order
approach) with those obtained by setting the torsion to zero in the
action before carrying out the variation (second-order approach). In
the latter case, the rotational gauge field is no longer an
independent field. In WGT (and also PGT and CGT), it may be written
explicitly in terms the other gauge fields, whereas in eWGT there
exists an implicit constraint relating all the gauge fields. In both
cases, one may arrive at simple expressions for the variational
derivatives in the second-order approach in terms of those from the
first-order approach.

The outline of this paper is as follows. In Section~\ref{sec:localsym}
we briefly review the concepts of local symmetries and dynamics in
classical field theory. We present our manifestly covariant
variational principle in Section~\ref{sec:mancovvp}, which is applied
to WGT and eWGT in Sections~\ref{sec:wgt} and \ref{sec:ewgt},
respectively. We conclude in Section~\ref{sec:conc}. In addition, in
Appendix~\ref{app:BHmethod}, we include a brief account of the
Bessel-Hagen method \cite{Bessel-Hagen21} for expressing the variation
of the vector potential in electromagnetism in a manifestly gauge-invariant
form; it is this approach that we generalise to gauge theories of
gravity in order to assist in directly obtaining manifestly covariant
conservation laws.

\section{Local symmetries and dynamics in classical field theory}
\label{sec:localsym}

We begin by presenting a brief outline of the consequences of local
symmetries for classical field theories, focusing in particular on
Noether's first and second theorems, the latter being discussed
surprisingly rarely in the literature. These considerations allow
one also to determine the dynamics of the fields.

Consider a spacetime manifold ${\cal M}$, labelled using some
arbitrary coordinates $x^\mu$, in which the dynamics of some set of
(tensor and/or spinor) fields $\chi(x) = \{\chi_A(x)\}$
($A=1,2,\ldots$) is described by the action\footnote{In our subsequent
  discussion, we will typically assume that ${\cal M}$ is Minkowski
  spacetime and $x^\mu$ are Cartesian inertial coordinates, but this
  is unnecessary for the analysis in this section.}
\begin{equation}
S = \int {\cal L}(\chi,\partial_\mu\chi,\partial_\mu\partial_\nu\chi)\,d^4x.
\label{eq:genmatteraction}
\end{equation}
It should be understood here that the label $A$ merely enumerates the
different fields, although (with some overloading of the notation) can
also be considered to represent one or more coordinate and/or local
Lorentz frame indices (either as subscripts or superscripts), which we
denote by lower case Greek and Roman letters,
respectively.
It is worth noting that, in general, each
field $\chi_A(x)$ may be either a matter field $\vpsi_A(x)$ or gauge
field $g_A(x)$. Allowing the Lagrangian density ${\cal L}$ in the
action (\ref{eq:genmatteraction}) to depend on field derivatives up to
second order is sufficient to accommodate all the gravitational gauge
theories that we will consider (and also general relativity).

Invariance of the action (\ref{eq:genmatteraction}) under the
infinitesimal coordinate transformation $x^{\prime\mu} = x^\mu +
\xi^\mu(x)$ and form variations $\delta_0\chi_A(x)$ in the fields
(where, importantly, the latter need not result solely from the coordinate
transformation)\footnote{Adopting Kibble's original notation, for an infinitesimal coordinate
  transformation $x^{\prime\mu} = x^\mu + \xi^\mu(x)$, the `form'
  variation $\delta_0\chi(x)\equiv\chi'(x)-\chi(x)$ is related to
  the `total' variation $\delta\chi(x)\equiv\chi'(x')-\chi(x)$ by
  $\delta_0\chi(x) = \delta\chi(x) - \xi^\mu\partial_\mu\chi(x)$.},
implies that
\be
\delta S  =  \int \left[\delta_0 {\cal L} + \partial_\mu(\xi^\mu {\cal L})\right]\,d^4x = 0,
\label{eq:genactioninv}
\ee
in which the form variation of the Lagrangian density is given by
\bea
\delta_0 {\cal L} \! = \! \pd{{\cal L}}{\chi_A}\delta_0\chi_A + 
\pd{{\cal L}}{(\partial_\mu\chi_A)}\delta_0(\partial_\mu\chi_A) +
\pd{{\cal L}}{(\partial_\mu\partial_\nu\chi_A)}\delta_0(\partial_\mu\partial_\nu\chi_A).
\label{eq:lagformvar}
\eea
One should note that $\delta_0$ commutes with partial derivatives and,
according to the usual summation convention, there is an implied sum
on the label $A$.  The integrand in the invariance condition
(\ref{eq:genactioninv}) can be rewritten directly using the product
rule to yield
\be
\delta S = \int \left(\frac{\delta {\cal L}}{\delta\chi_A}\delta_0\chi_A +
\partial_\mu J^\mu \right) \,d^4x = 0,
\label{eq:genactioninv2}
\ee
where the Euler--Lagrange variational derivative $\delta {\cal
  L}/\delta\chi_A$ and the Noether current $J^\mu$ are given,
respectively, by
\begin{subequations}
\bea
\frac{\delta {\cal L}}{\delta\chi_A}
& = & \pd{{\cal L}}{\chi_A}
- \partial_\mu\left(\pd{{\cal L}}{(\partial_\mu\chi_A)}\right)
+ \partial_\mu\partial_\nu\left(
\pd{{\cal L}}{(\partial_\mu\partial_\nu\chi_A)}
\right),\phantom{AA} \label{eq:eleq}\\
J^\mu & = & 
\left[\momc{\mu}{A}
-\partial_\nu\left(\pd{{\cal L}}{(\partial_\mu\partial_\nu\chi_A)}\right)
\right]\delta_0\chi_A + 
\pd{\cal L}{(\partial_\mu\partial_\nu\chi_A)}\partial_\nu(\delta_0\chi_A)
+\xi^\mu
{\cal L}.
\label{eq:noethercurrent}
\eea
\end{subequations}
%

It is worth noting that the equations of motion for the fields
$\chi_A(x)$ are also obtained by considering the behaviour of the
action under variations of the fields, but with the coordinate system
kept fixed, so that $\xi^\mu(x)=0$. One further assumes that the
variations $\delta_0\chi_A(x)$ vanish on the boundary of the
integration region of the action, and also that their first derivatives
$\partial_\mu(\delta_0\chi_A(x))$ vanish in the case where ${\cal
  L}$ contains second derivatives of the fields. In order for the
action to be stationary $\delta S = 0$ with respect to arbitrary such
variations $\delta_0\chi_A(x)$ of the fields, one thus requires
(\ref{eq:genactioninv2}) to hold in these circumstances, which
immediately yields the equations of motion $\delta {\cal
  L}/\delta\chi_A = 0$.

Returning to considering (\ref{eq:genactioninv2}) as denoting the
invariance of the action (\ref{eq:genmatteraction}) under some
general infinitesimal coordinate transformation $x^{\prime\mu} = x^\mu +
\xi^\mu(x)$ and form variations $\delta_0\chi_A(x)$ in the fields
(which need not vanish on the boundary of the integration region), one
sees that if the field equations $\delta {\cal L}/\delta\chi_A = 0$
are satisfied for all the fields, then (\ref{eq:genactioninv2})
reduces to the (on-shell)\footnote{We use Dirac's notation $F\eom 0$
  for local functions $F$ that vanish on-shell (or {\it weakly}
  vanish), i.e.\ when the equations of motion $\delta {\cal
    L}/\delta\chi_A = 0$ are satisfied for {\it all} the fields. We
  further denote by $F\eomm 0$ and $F\eomg 0$ when functions vanish if
  only the equations of motion of the matter or gauge fields,
  respectively, need be satisfied.} `conservation law' $\partial_\mu
J^\mu \eom 0$, which holds up to a total divergence of any quantity
that vanishes on the boundary of the integration region of the action
(\ref{eq:genmatteraction}). This is the content of Noether's first
theorem, which applies both to global and local symmetries.

We will focus on the invariance of the action
(\ref{eq:genmatteraction}) under a local symmetry.  In particular, we
consider the (usual) case in which the form variations of the fields
can be written as
\be 
\delta_0\chi_A = \lambda^C f_{AC}(\chi,\partial\chi) +
(\partial_\mu\lambda^C) f_{AC}^\mu(\chi,\partial\chi),
\label{eq:localformvar}
\ee
where $\lambda^C=\lambda^C(x)$ are a collection of independent
arbitrary functions of spacetime position, enumerated by the label $C$, and
$f_{AC}(\chi,\partial\chi)$ and $f_{AC}^\mu(\chi,\partial\chi)$ are
two collections of given functions that, in general, may depend on all
the fields and their first derivatives.  The general form
(\ref{eq:localformvar}) usually applies only when $\chi_A = g_A$ is a
gauge field, whereas typically $f_{AC}^\mu(\chi,\partial\chi) = 0$ if
$\chi_A = \vpsi_A$ is a matter field. For each value of $C$, the
function $\lambda^C(x)$ represents a set of infinitesimal functions
carrying one or more coordinate or local Lorentz frame indices.  It is
worth noting that on substituting (\ref{eq:localformvar}) into
(\ref{eq:noethercurrent}), one obtains an expression for the current
$J^\mu$ where the first term is proportional to
(\ref{eq:localformvar}) and, in the event that ${\cal L}$ depends on
second derivatives of the fields, the second term is proportional to
the first derivative of (\ref{eq:localformvar}), which itself contains
second derivatives of the functions $\lambda^C(x)$.


Using the expression (\ref{eq:localformvar}), and again employing 
the product rule, the corresponding variation of the action
(\ref{eq:genactioninv2}) is given by (suppressing functional
dependencies for brevity)
\be
\delta S \! = \! \int \lambda^C \left[f_{AC}\frac{\delta{\cal
      L}}{\delta\chi_A}-\partial_\mu\left(f_{AC}^\mu\frac{\delta{\cal
      L}}{\delta\chi_A}\right)\right] + \partial_\mu (J^\mu -
S^\mu)\,d^4x \! = \! 0,
\ee
where we define the new current $S^\mu \equiv -\lambda^C f_{AC}^\mu
\delta{\cal L}/\delta\chi_A$. It is worth noting that $S^\mu$ depends
much more simply than $J^\mu$ on the functions $\lambda^C$.
Since the $\lambda^C$ are
arbitrary functions, for the action to be invariant one requires
the separate conditions
\begin{subequations}
\label{eq:n2cond}
\bea
f_{AC}\frac{\delta{\cal
      L}}{\delta\chi_A}-\partial_\mu\left(f_{AC}^\mu\frac{\delta{\cal
      L}}{\delta\chi_A}\right) &=& 0,\label{eq:n2cond1}\\
\partial_\mu(J^\mu -S^\mu) & = & 0,\label{eq:n2cond2}
\eea
\end{subequations}
where the former hold for each value of $C$ separately and the latter
holds up to a total divergence of a quantity that vanishes on the
boundary of the integration region.

The first set of conditions (\ref{eq:n2cond1}) are usually interpreted
as conservation laws, which are covariant under the local symmetry,
although not manifestly so in the form given above. The condition
(\ref{eq:n2cond2}) implies that $J^\mu = S^\mu + \partial_\nu
Q^{\nu\mu}$, where $Q^{\nu\mu} = -Q^{\mu\nu}$, so the two currents
coincide up to a total divergence, which is notable given their very
different dependencies on the functions $\lambda^C$, $f_{AC}$ and
$f_{AC}^\mu$, as described above. By contrast with the case of a
global symmetry,\footnote{For a
  global symmetry, the $\lambda^C$ are constants and so the second
  term on the RHS of (\ref{eq:localformvar}) vanishes. The Noether current
  (\ref{eq:noethercurrent}) can be then written as
\be
J^\mu  =  \lambda^C\left\{
\left[\momc{\mu}{A}
-\partial_\nu\left(\pd{{\cal L}}{(\partial_\mu\partial_\nu\chi_A)}\right)
\right]f_{AC} + 
\pd{\cal L}{(\partial_\mu\partial_\nu\chi_A)}\partial_\nu f_{AC}
+\xi_C^\mu{\cal L}\right\} \equiv \lambda^C J_C^\mu,\nonumber
\ee
where $\xi_C^\mu$ are a given set of functions such that 
$\xi^\mu=\lambda^C\xi_C^\mu$, and we have also defined
the further set of functions $J_C^\mu$. One can then replace the two conditions
(\ref{eq:n2cond}) with the following single condition that is
not satisfied identically on-shell:
\be
f_{AC}\frac{\delta{\cal L}}{\delta\chi_A}-\partial_\mu J_C^\mu = 0.
\nonumber
\ee
} if the field equations $\delta{\cal L}/\delta\chi_A = 0$ are
satisfied for all fields, then the conservation laws
(\ref{eq:n2cond1}) hold identically and the new current vanishes
$S^\mu \eom 0$, so that $J^\mu \eom \partial_\nu Q^{\nu\mu}$.  Thus,
the conditions (\ref{eq:n2cond1}--\ref{eq:n2cond2}) effectively
contain no information on-shell, which is essentially the content of
Noether's second theorem \cite{Avery16}.

Nonetheless, the on-shell condition that {\it all} the field equations
$\delta{\cal L}/\delta\chi_A = 0$ are satisfied can only be imposed if
${\cal L}$ is the {\it total} Lagrangian density, and not if ${\cal
  L}$ corresponds only to some {\it subset} thereof (albeit one for
which the corresponding action should still be invariant under the
local symmetry).  In particular, suppose one is considering a field
theory for which the total Lagrangian density ${\cal L}_{\rm T} =
{\cal L}_{\rm M} + {\cal L}_{\rm G}$, where ${\cal L}_{\rm G}$
contains every term that depends {\it only} on the gauge fields $g_A$
and/or their derivatives, and ${\cal L}_{\rm M}$ contains all the
remaining terms. Thus, if ${\cal L} = {\cal L}_{\rm M}$, then only the
matter field equations $\delta{\cal L}/\delta\vpsi_A = 0$ can be
imposed, whereas if ${\cal L} = {\cal L}_{\rm G}$ none of the field
equations can be imposed.  In either case, the surviving terms in
(\ref{eq:n2cond1}--\ref{eq:n2cond2}) do contain information
\cite{Forger04}.

\section{Manifestly covariant variational principle}
\label{sec:mancovvp}

In the standard variational approach outlined above, one sees
immediately from the plethora of partial derivatives throughout the
analysis that the various expressions obtained are not, in general,
manifestly covariant under the symmetry group to which the action is
invariant.  In particular, although the equations of motion
$\delta{\cal L}/\delta\chi_A = 0$ for each field must be covariant
under this symmetry group, it is clear that those derived from
(\ref{eq:eleq}) are not manifestly so. Moreover, the conservation laws
(\ref{eq:n2cond1}) suffer from the same shortcoming, but must also be
expressible in a manifestly covariant form. By contrast, the currents
$J^\mu$ and $S^\mu$ are not covariant (manifestly or otherwise), in
general, since they both contain the arbitrary functions
$\lambda^C(x)$, and $J^\mu$ also contains their partial
derivatives. To obtain manifestly covariant variational derivatives
and conservation laws directly, it is expedient to take a different
approach that begins afresh by reconsidering the variation of the
action in (\ref{eq:genactioninv}).

We are primarily concerned here with gauge theories of
gravity. In constructing such theories, one typically begins with an
action dependent only on some set of matter fields $\vpsi_A$, which is
defined on Minkowski spacetime ${\cal M}$ in Cartesian inertial
coordinates $x^\mu$ (which we will assume henceforth), and is
invariant under some global spacetime symmetry group $\pazocal{G}$,
where the coefficients $\lambda^C$ in (\ref{eq:localformvar}) are
constants. One then gauges the group $\pazocal{G}$ by demanding that
the action be invariant with respect to (infinitesimal, passively
interpreted) general coordinate transformations (GCTs) and the local
action of the subgroup $\pazocal{H}$ (say), obtained by setting the
translation parameters of $\pazocal{G}$ to zero (which leaves the
origin invariant), and allowing the remaining group parameters to
become independent arbitrary functions of position.  For example, if
one considers global Weyl invariance, then
$\{\lambda^1,\lambda^2,\lambda^3\} =
\{a^\alpha,\omega^{\alpha\beta},\rho\}$, which denote a global
spacetime translation, rotation and dilation, respectively.  The
symmetry is then `promoted' to a local one by allowing $\lambda^C(x)$
to become arbitrary functions of spacetime position $x$. For local
Weyl invariance, one thus has
$\{\lambda^1(x),\lambda^2(x),\lambda^3(x)\}
=\{a^\alpha(x),\omega^{ab}(x),\rho(x)\}$, where $a^\alpha(x)$ is
interpreted as an infinitesimal general coordinate transformation and
is usually denoted instead by $\xi^\alpha(x)$, and $\omega^{ab}(x)$ and
$\rho(x)$ denote a position-dependent rotation of the local Lorentz
frames and a position-dependent dilation, respectively.  For the
action to remain invariant under the localised symmetry necessitates
the introduction of gravitational gauge fields $g_A$ with prescribed
transformation properties under the action of the localised symmetry.
We will also maintain the somewhat unorthodox viewpoint, albeit hinted
at in Kibble's original paper, of considering the gravitational gauge
fields as fields in Minkowski spacetime, without attaching any
geometric interpretation to them. Consequently, we will adopt a global
Cartesian inertial coordinate system $x^\mu$ in our Minkowski
spacetime, which greatly simplifies calculations, but more general
coordinate systems may be straightforwardly accommodated, if required
\cite{eWGTpaper}.

For an action (\ref{eq:genmatteraction}) containing both matter fields
$\chi_A = \psi_A$ and gauge fields $\chi_A = g_A$ to be invariant
under a local symmetry of the form (\ref{eq:localformvar}), one
requires the Lagrangian density ${\cal L}$ to be covariant under this
symmetry. One typically always requires invariance of the action under
at least (infinitesimal) general coordinate transformations (GCTs),
which can be considered as promoting the set of constants $\lambda^C$
representing global translations to arbitrary functions of position;
this necessitates the introduction of the corresponding translational
gravitational gauge field, which we will denote by ${h_a}^\mu$ and its
inverse by ${b^a}_\mu$ (such that ${h_a}^\mu {b^a}_\nu =
\delta^\mu_\nu$ and ${h_a}^\mu {b^c}_\mu = \delta_a^c$).  It is
therefore convenient to write the Lagrangian density as the product
${\cal L} = h^{-1}L$, where $h = {\rm det}({h_a}^\mu)$ is a scalar density,
since $h^{-1}\, d^4x$ is an invariant volume element under
GCTs.\footnote{We will also denote  $h^{-1}$ by $b$ where $b \equiv
  \mbox{det}({b^a}_\mu)$.} The remaining factor $L$, which we term the
Lagrangian, is also a scalar density constructed from covariant
quantities.\footnote{It should be noted that if the set of local
  symmetries (\ref{eq:localformvar}) of the action include local scale
  transformations, then the Weyl weights of the scalar densities $b$
  and $L$ should sum to zero, namely $w(b) + w(L) = 0$, so that the
  action $S$ is invariant.}  These typically include the matter fields
$\vpsi_A$ themselves and their covariant derivatives, together with
the field strength tensors ${\cal F}_B$ of the gauge fields $g_B$,
which typically depend both on the gauge fields themselves and their
partial derivatives (where we have adopted a `symbolic' form that suppresses coordinate and
local Lorentz frame indices).  In this section, we will denote the
generic covariant dervative by $\Db_a \equiv {h_a}^\mu \mathsf{D}_\mu
= {h_a}^\mu(\partial_\mu + \Gamma_\mu)$, where $\Gamma_\mu$ is a
linear combination of the generators of the subgroup $\pazocal{H}$
that may depend, in general, on the gauge fields $g_A$ and their first
derivatives $\partial g_A$ (note that we will occasionally retain the
indices on covariant derivatives, when convenient to do so).  In any
case, one can thus denote the functional dependence of the Lagrangian
symbolically by $L = L(\vpsi_A,\Db_a\vpsi_A,{\cal F}_B)$.

\subsection{Manifestly covariant variational derivatives}

We begin by rewriting the variation of the action
(\ref{eq:genactioninv}) so that one can directly identify manifestly
covariant forms for the variational derivatives $\delta{\cal
  L}/\delta\chi_A$.  One must first obtain a covariant form for the
divergence in (\ref{eq:genactioninv}) by constructing a further
covariant derivative operator $\mathfrak{D}_a$ such that, for any
coordinate vector $V^\mu$ (of the same Weyl weight as the Lagrangian
density ${\cal L}$), one has $\partial_\mu V^\mu =
h^{-1}\mathfrak{D}_a(h{\cal V}^a)$, where we define the local Lorentz
frame vector\footnote{We will typically denote a quantity possessing
only Roman indices (and its contractions over such indices) as the
calligraphic font version of the kernel letter of the corresponding
quantity possessing only Greek indices (following \cite{eWGTpaper}),
with the exception of quantities having Greek or lower-case kernel
letters.}  ${\cal V}^a = {b^a}_\mu V^\mu$. The construction of such an
operator requires one first to define the field strength tensor
${\pazocal{T}^a}_{bc} = 2{h_b}^{\mu}{h_c}^{\nu} \mathsf{D}_{[\mu}
  {b^a}_{\nu]}$ of the translational gauge field, which has the unique
(up to a sign) non-trivial contraction $\pazocal{T}_b \equiv
{\pazocal{T}^a}_{ba} = h\mathsf{D}_{\mu}(h^{-1}{h_b}^\mu)$. It is then
straighforward to show that the required derivative operator is given
by $\mathfrak{D}_a = \Db_a + \Tb_a$.

One may then rewrite the variation of the action
(\ref{eq:genactioninv}) in the alternative form
\be
\delta S  =  \int \left[\delta_0 {\cal L} 
+ h^{-1}(\Db_a + \Tb_a)(\xi^a L)\right]\,d^4x = 0,
\label{eq:genactioninv-mancov}
\ee
in which $\xi^a = {b^a}_\mu\xi^\mu$ and the form variation of the Lagrangian density
(\ref{eq:lagformvar}) can be rewritten symbolically as
\be
\delta_0{\cal L} = h^{-1}\left[\bpd{L}{\vpsi_A}\,\delta_0\vpsi_A + \pd{L}{(\Db_a\vpsi_A)}\,\delta_0(\Db_a\vpsi_A) + 
\pd{L}{{\cal F}_B}\delta_0{\cal F}_B \right] + L\,\delta_0 h^{-1},
\label{eq:lagformvar2}
\ee
where $\bar{\partial}L/\partial\vpsi \equiv [\partial L(\vpsi,\Db_a
  u,\ldots)/\partial \vpsi]_{u=\vpsi}$, so that $\vpsi$ and $\Db_a
\vpsi$ are treated as independent variables, rather than $\vpsi$ and
$\partial_\mu\vpsi$.  In order to progress further, the variations
$\delta_0(\Db_a\vpsi_A)$, $\delta_0{\cal F}_B$ and $\delta_0h^{-1}$ in
(\ref{eq:lagformvar2}) must be expressed in terms of the variations
$\delta_0\vpsi_A$ and $\delta_0 g_B$, respectively, of the matter and
gauge fields themselves. In so doing, one typically encounters terms
of the (symbolic) form $\Db(\delta_0\vpsi_A)\,\partial
L/\partial(\Db\vpsi_A)$ and $\Db(\delta_0 g_{B})\,\partial
L/\partial{\cal F}_B$, which can be accommodated by considering the
quantity $(\Db_a+\Tb_a) (h{\cal V^a})$, where (again in symbolic form)
$h{\cal V} \sim \delta_0\vpsi_A \,\partial L/\partial(\Db\vpsi_A) +
\delta_0 g_B\, \partial L/\partial{\cal F}_B$, and then using the
product rule.  Following such a procedure, one may rewrite
(\ref{eq:lagformvar2}) in the general form
\be
\delta_0{\cal L} = h^{-1}\left[\alpha^A\,\delta_0\vpsi_A + \beta^B\,\delta_0 g_B
+ (\Db_a + \Tb_a) (h{\cal V}^a)\right],
\label{eq:lagformvar3}
\ee
where $\alpha^A$ and $\beta^B$ are manifestly covariant expressions
that typically depend on $\vpsi_A$, $\partial L/\partial\vpsi_A$ and
${\cal F}_B$, together with $\partial L/\partial(\Db\vpsi_A)$ and
$\partial L/\partial{\cal F}_B$ and their covariant
derivatives. Inserting (\ref{eq:lagformvar3}) into
(\ref{eq:genactioninv-mancov}), Noether's first theorem
(\ref{eq:genactioninv2})
becomes
\be
\delta S  =  \int \left[\alpha^A\,\delta_0\vpsi_A + \beta^B\,\delta_0 g_B
+ (\Db_a + \Tb_a)(h{\cal J}^a)\right]\,h^{-1}\,d^4x = 0,
\label{eq:genactioninv-mancov2}
\ee
where the current $h{\cal J}^a = h{\cal V}^a + \xi^a L$ has
the symbolic form
\be
h{\cal J} \sim \pd{L}{(\Db\vpsi_A)}\,\delta_0\vpsi_A + 
\pd{L}{{\cal F}_B}\,\delta_0 g_B + \xi L.
\label{eq:jsymbolic}
\ee
By comparing (\ref{eq:genactioninv2}) and
(\ref{eq:genactioninv-mancov2}), and noting that $h^{-1}(\Db_a + \Tb_a)(h{\cal J}^a) = \partial_\mu J^\mu$, one may then
immediately identify manifestly covariant expressions for the
variational derivatives with respect to the matter and gauge fields,
respectively, as
\be
\frac{\delta {\cal L}}{\delta\vpsi_A} = b\alpha^A,\qquad
\frac{\delta {\cal L}}{\delta g_B} = b\beta^B.
\label{eq:varderivs-mancov}
\ee
If one does not wish to distinguish between matter and gauge fields,
one can instead denote the above relations generically by $\delta{\cal
  L}/\delta\chi_A = b\gamma^A$, where $\gamma^A$ is a manifestly
covariant expression.

\subsection{Manifestly covariant conservation laws}

We now turn to the direct construction of manifestly covariant
expressions for the conservation laws (\ref{eq:n2cond1}).  Clearly,
the manifestly covariant expressions (\ref{eq:varderivs-mancov}) may
now be used for the variational derivatives, but one encounters two
remaining issues, namely the presence of the explicit partial
derivative in the second term in (\ref{eq:n2cond1}), and the fact that
the functions $f_{AC}$ and $f^\mu_{AC}$ may not be covariant
quantities. Indeed, the latter problem always occurs when the
functions $\lambda^C(x)$ (say for $C=1$) correspond to GCTs, such that
$\lambda^1(x) = \{\xi^\alpha(x)\}$; this arises because
$\delta_0\chi_A = \delta\chi_A - \xi^\alpha \partial_\alpha\chi_A$ for
any field and so $f_{A1}$ always contains the non-covariant term
$-\partial_\alpha\chi_A$. Other functions from the sets $f_{AC}$ and
$f^\mu_{AC}$ may also be non-covariant, depending on the gauge theory
under consideration. 

Nonetheless, it is important to recall that the conservation law
(\ref{eq:n2cond1}) holds for {\it any} set of form variations of the
fields (\ref{eq:localformvar}) that leave the action invariant. In
particular, by generalising the approach first proposed by
Bessel-Hagen for electromagnetism (see Appendix~\ref{app:BHmethod}),
one can choose specific forms for the functions $\lambda^C(x)$ for $C
\neq 1$ in terms of $\lambda^1(x)$ and the non-translational gauge
fields $g_B$, such that all the functions $f^\mu_{AC}$ become
(manifestly) covariant (as typically do many of the functions
$f_{AC}$). In this case, one may then write the second term in
(\ref{eq:n2cond1}) by extending the definition of the covariant
derivative $(\Db_a + \Tb_a)$ to accommodate any additional free indices
represented by the subscript $C$. In particular, it is convenient to
require that for any quantity ${V_C}^\mu$ with this index structure
(and the same Weyl weight as the Lagrangian density ${\cal L}$), one
has $h^{-1}(\Db_a + \Tb_a)(h {b^a}_\mu {V_C}^\mu) = \mathsf{D}_\mu
{V_C}^\mu = (\partial_\mu + \Gamma_\mu){V_C}^\mu$, so that in the case
where $C$ does not represent any additional indices one recovers the
original requirement that $h^{-1}(\Db_a + \Tb_a)(h {b^a}_\mu V^\mu) =
\partial_\mu V^\mu $. One may then write the conservation law
(\ref{eq:n2cond1}) as
\be 
(\Db_a + \Tb_a)({b^a}_\mu f_{AC}^\mu\gamma^A) 
-(f_{AC}+\Gamma_\mu f_{AC}^\mu)\gamma^A = 0.\label{eq:n2cond1-mancov} 
\ee
The first term on the LHS of (\ref{eq:n2cond1-mancov}) is now
manifestly covariant. Consequently, although the second term on the
LHS is not manifestly covariant, it must also be expressible in such a
form; indeed, one typically finds that this second term immediately
assembles as such, as we will demonstrate in Sections~\ref{sec:wgt}
and \ref{sec:ewgt} where we apply this approach to WGT and eWGT,
respectively.

\subsection{Relationship between currents in Noether's second theorem}
\label{subsec:wgtnt2}

Finally, we consider the relationship (\ref{eq:n2cond2}) between the
two currents $J^\mu$ and $S^\mu$. As noted above, both currents depend
on the functions $\lambda^C$ and so neither is covariant. Nonetheless,
from the above discussion, one may rewrite (\ref{eq:n2cond2}) as
$(\Db_a + \Tb_a) [h({\cal J}^a -{\cal S}^a)] = 0$, in which
\be h{\cal S}^a = -\lambda^C h {b^a}_\mu
f_{AC}^\mu\frac{\delta{\cal L}}{\delta\chi_A} = -\lambda^C {b^a}_\mu
f_{AC}^\mu\gamma^A = -\lambda^C {b^a}_\mu
(f_{AC}^\mu\alpha^A+f_{BC}^\mu\beta^B),
\label{eq:generalscurrent}  
\ee
where we have used the relations (\ref{eq:varderivs-mancov}) to write
the final expression in terms of the matter fields and gauge fields
separately, in keeping with the (symbolic) expression
(\ref{eq:jsymbolic}) for $h{\cal J}^a$.  Thus, $h{\cal S}^a$ has the
form of linear combination of terms that are manifestly covariant (or
can be made so using a generalisation of the Bessel-Hagen method) with
coefficients $\lambda^C$. Turning to $h{\cal J}^a$, if one substitutes
(\ref{eq:localformvar}) into (\ref{eq:jsymbolic}), and recalls that
$f^\mu_{AC}$ typically vanishes for matter fields, one obtains the
(symbolic) expression
\be
h{\cal J} \sim \lambda^C\left(\pd{L}{(\Db\vpsi_A)} f_{AC} +
\delta_C^1 L\right)
+ \pd{L}{{\cal F}_B}(f_{BC}\lambda^C + f^\mu_{BC}\partial_\mu\lambda^C),
\ee
where we have again assumed that $C=1$ corresponds to GCTs. One may
show, in general, that the forms of the manifestly covariant
expressions $\alpha^A$ and $\beta^B$ obtained in
(\ref{eq:lagformvar3}) guarantee that the relationship
$(\Db_a + \Tb_a)[h({\cal J}^a -{\cal S}^a)] = 0$ is
satisfied, and so it contains no further information. It is worth
noting, however, that for the special case in which $L$ does not
depend on the gauge field strengths, such that $\partial L/\partial
{\cal F}_B = 0$, the relationship takes the form
\be
(\Db_a + \Tb_a)\left[
\lambda^C
\left(\pd{L}{(\Db_a\vpsi_A)}f_{AC}+\delta_C^1 L + {b^a}_\mu f^\mu_{AC}\alpha^A
\right)
\right] = 0,
\ee
which may be satisfied by requiring the term in parentheses to vanish
for each value of $C$. In so doing, one obtains a straightforward
expression for $\alpha^A$, which one can show agrees with that
obtained in (\ref{eq:lagformvar3}).

\medskip
The procedures presented in this section are best illustrated by
example and we apply them to WGT and eWGT in Sections~\ref{sec:wgt}
and \ref{sec:ewgt}, respectively. As we will also show in these
examples, the general approach outlined above further lends itself to
elucidating the relationship between first- and second-order
variational derivatives.

\section{Weyl gauge theory}
\label{sec:wgt}

For WGT, the Lagrangian density has the usual form ${\cal L} = h^{-1}L$,
where the translational gauge field ${h_a}^\mu$ is assigned a Weyl
weight $w=-1$, so that 
$h = {\rm det}({h_a}^\mu)$ and $L$ are scalar densities both of Weyl
weight $w=-4$, and hence the action $S$ is
invariant under local scale transformations.  The Lagrangian has the
functional dependencies
\be
L = L(\vpsi_A, {\cal D}^\ast_a\vpsi_A,{\cal R}_{abcd},{\cal
  T}^\ast_{abc},{\cal H}_{ab}),
\label{eq:wgtlag}
\ee
where $\vpsi_A$ are the matter fields, which typically include a
scalar compensator field of Weyl weight $w=-1$ (that we sometimes
denote also by $\phi$), and their covariant derivatives are denoted in
this section by \cite{Blagojevic02,eWGTpaper,CGTpaper}
\be
{\cal D}^\ast_a\vpsi_A = {h_a}^\mu D^\ast_\mu \vpsi_A
= {h_a}^\mu (\partial_\mu + \Gamma^\ast_\mu)\vpsi_A = {h_a}^\mu (\partial_\mu +
\tfrac{1}{2}{A^{cd}}_\mu\Sigma_{cd}+w_AB_\mu)\vpsi_A,
\label{eq:wgtcovderivdef}
\ee
in which ${h_a}^\mu$ (with inverse ${b_a}^\mu$), ${A^{ab}}_\mu =
-{A^{ba}}_\mu$ and $B_\mu$ are the translational, rotational and
dilational gravitational gauge fields, respectively, and $\Sigma_{ab}
= -\Sigma_{ba}$ are the generator matrices of the $\mbox{SL}(2,C)$
representation to which the field $\vpsi_A$ belongs.\footnote{The
  asterisks in the definition of the derivative operator are intended
  simply to distinguish it from the usual notation used
  \cite{Blagojevic02,eWGTpaper,CGTpaper} for the covariant derivative ${\cal
    D}_a\vpsi_A = {h_a}^\mu D_\mu \vpsi_A = {h_a}^\mu (\partial_\mu +
  \Gamma_\mu)\vpsi_A = {h_a}^\mu (\partial_\mu +
  \tfrac{1}{2}{A^{cd}}_\mu\Sigma_{cd})\vpsi_A$ of Poincar\'e gauge
  theory (PGT), and should not be confused with the operation of
  complex conjugation.} In the expression (\ref{eq:wgtcovderivdef}),
each field is assumed to have weight $w_A$ (note that this appearance
of the index $A$ is purely a label and hence is understood never to be
summed over). It is also convenient to define the further derivative
operator $\partial^\ast_\mu\vpsi_A = (\partial_\mu + w_A
B_\mu)\vpsi_A$, of which we will make occasional use.

Under infinitesimal local Weyl transformations consisting of GCTs,
rotations of the local Lorentz frames and dilations, which are
parameterised by $\xi^\mu(x)$, $\omega^{ab}(x)$ and $\rho(x)$,
respectively, a matter field $\vpsi$ of weight $w$ and the gauge fields
transform as \cite{Blagojevic02,CGTpaper}
\begin{subequations}
\label{eq:wgtfieldtrans}
\bea
\delta_0\vpsi & = & -\xi^\nu\partial_\nu\vpsi +
(\tfrac{1}{2}\omega^{ab}\Sigma_{ab} + w\rho)\vpsi,
\label{eq:vpsitrans}\\
\delta_0{h_a}^\mu & = & 
-\xi^\nu\partial_\nu {h_a}^\mu + {h_a}^\nu\partial_\nu\xi^\mu
- ({\omega^b}_{a} +\rho\,\delta^b_a){h_b}^\mu,\phantom{AAA} 
\label{eq:htrans}\\
\delta_0 {A^{ab}}_\mu & = &-\xi^\nu\partial_\nu{A^{ab}}_\mu
-{A^{ab}}_\nu \partial_\mu\xi^\nu
-2{{\omega^{[a}}_{c}A^{b]c}}_\mu 
-\partial_\mu\omega^{ab}, \label{eq:atrans}\\
\delta_0 B_\mu & = &  -\xi^\nu\partial_\nu B_\mu -B_\nu\partial_\mu\xi^\nu
-\partial_\mu\rho,\label{eq:btrans}
\eea
\end{subequations}
from which one may verify that ${\cal D}^\ast_a\vpsi_A$ does indeed
transform covariantly under (infinitesimal) local Weyl transformations
with weight $w-1$ \cite{eWGTpaper,CGTpaper}.

The remaining quantities ${\cal R}_{abcd}$, ${\cal T}^\ast_{abc}$,
${\cal H}_{ab}$ in (\ref{eq:wgtlag}) are the field strength tensors of
the rotational, translational and dilational gauge fields,
respectively, which are defined through the action of the commutator
of two covariant derivatives on some field $\vpsi$ of weight $w$ by
\be
[{\cal D}^\ast_c,{\cal D}^\ast_d]\vpsi =
(\tfrac{1}{2}{{\cal R}^{ab}}_{cd}\Sigma_{ab} + w{\cal H}_{cd}
- {{\cal T^\ast}^a}_{cd}{\cal D}^\ast_a)\vpsi.
\label{eq:dcommwgt}
\ee
The field strengths have the forms ${{\cal R}^{ab}}_{cd}
= {h_a}^{\mu}{h_b}^{\nu}{R^{ab}}_{\mu\nu}$, ${\cal H}_{cd} =
       {h_c}^\mu {h_d}^\nu H_{\mu\nu}$ and ${{\cal T}^{\ast a}}_{bc} =  {h_b}^{\mu}{h_c}^{\nu} {T^{\ast a}}_{\mu\nu}$, where
\begin{subequations}
\bea
{R^{ab}}_{\mu\nu}  & = &  2(\partial_{[\mu} {A^{ab}}_{\nu]} +
\eta_{cd}{A^{ac}}_{[\mu}{A^{db}}_{\nu]}),
\label{rfsdef} \\
H_{\mu\nu} & = & 2\partial_{[\mu} B_{\nu]},
\label{dfsdef}\\
{T^{\ast a}}_{\mu\nu} & = & 2D^\ast_{[\mu} {b^a}_{\nu]}.
\label{tfsdef}
\eea
\end{subequations}
From the transformation laws (\ref{eq:wgtfieldtrans}), it is
straightforward to verify that, in accordance with their index
structures, the gauge field strength tensors ${{\cal R}^{ab}}_{cd}$,
${\cal H}_{cd}$ and ${{\cal T}^{\ast a}}_{bc}$ are invariant under
GCTs, and transform covariantly under local Lorentz transformations
and dilations with Weyl weights $w = -2$, $w=-2$ and $w=-1$,
respectively \cite{eWGTpaper,CGTpaper}.

It is worth noting that ${{\cal R}^{ab}}_{cd}$ has the same functional
form as the rotational field strength in PGT, but that ${{\cal
    T}^{\ast a}}_{bc} = {{\cal T}^a}_{bc} + \delta^a_c{\cal B}_b -
\delta^a_b{\cal B}_c$, where ${{\cal T}^a}_{bc}$ is the translational
field strength in PGT; we also define ${\cal B}_a = {h_a}^\mu B_\mu$.
Moreover, using the expression (\ref{tfsdef}) and defining the
quantities ${c^{\ast a}}_{bc} \equiv 2{h_b}^\mu {h_c}^\nu
\partial^\ast_{[\mu} {b^a}_{\nu]}$, one may show that the fully
anholonomic rotational gauge field ${{\cal A}^{ab}}_c \equiv {h_c}^\mu
{A^{ab}}_\mu$ can be written as \cite{Blagojevic02,eWGTpaper}
\be
{\cal A}_{abc} = \tfrac{1}{2}(c^\ast_{abc}+c^\ast_{bca}-c^\ast_{cab})
-\tfrac{1}{2}({\cal T}^\ast_{abc}+{\cal T}^\ast_{bca}-{\cal
  T}^\ast_{cab}).
\label{afromht}
\ee

It is also convenient for our later development to obtain the Bianchi
identities satisfied by the gravitational gauge field strengths
${{\cal R}^{ab}}_{cd}$, ${{\cal T}^{\ast a}}_{bc}$ and ${\cal H}_{ab}$
in WGT. These may be straightforwardly derived from the Jacobi
identity applied to the generalised covariant derivative, namely
$[{\cal D}^\ast_a,[{\cal D}^\ast_b,{\cal D}^\ast_c]]\vpsi + [{\cal
    D}^\ast_c,[{\cal D}^\ast_a,{\cal D}^\ast_b]]\vpsi + [{\cal
    D}^\ast_b,[{\cal D}^\ast_c,{\cal D}^\ast_a]]\vpsi =0$.  Inserting
the form (\ref{eq:wgtcovderivdef}) for the WGT generalised covariant
derivative, one quickly finds the three Bianchi identities
\cite{eWGTpaper}\footnote{Note that these expressions correct a
  typographical error in \cite{eWGTpaper} by reversing the sign of each
  term containing ${\cal H}_{ab}$.}
\begin{subequations}
\begin{eqnarray}
{\cal D}^\ast_{[a}{{\cal R}^{de}}_{bc]}-{{\cal T}^{\ast
    f}}_{[ab} {{\cal R}^{de}}_{c]f} & = & 0, \label{wgtbi1} \\
{\cal D}^\ast_{[a}{{\cal T}^{\ast d}}_{bc]}-{{\cal T}^{\ast
    e}}_{[ab} {{\cal T}^{\ast d}}_{c]e}-{{\cal R}^{d}}_{[abc]} - {\cal H}_{[ab}\delta^d_{c]}
 & = & 0, \label{wgtbi2} \\
{\cal D}^\ast_{[a}{{\cal H}}_{bc]} -{{\cal T}^{\ast
    e}}_{[ab} {{\cal H}}_{c]e} & = & 0. \label{wgtbi3}
\end{eqnarray}
\end{subequations}
By contracting over various indices, one also obtains the
following non-trivial contracted Bianchi identities:
\begin{subequations}
\begin{eqnarray}
{\cal D}^\ast_{a}{{\cal R}^{ae}}_{bc}-2
{\cal D}^\ast_{[b}{{\cal R}^{e}}_{c]}
-2{{\cal T}^{\ast
    f}}_{a[b} {{\cal R}^{ae}}_{c]f}
-{{\cal T}^{\ast
    f}}_{bc} {{\cal R}^{e}}_{f}
 & = &   0, \label{wgtcbi1} \\
{\cal D}^\ast_{a}({{\cal R}^{a}}_{c}
-\tfrac{1}{2}\delta^a_c{\cal R})
+{{\cal T}^{\ast
    f}}_{bc} {{\cal R}^{b}}_{f}
+\tfrac{1}{2}{{\cal T}^{\ast
    f}}_{ab} {{\cal R}^{ab}}_{cf}
& = &  0, \label{wgtcbi2} \\
{\cal D}^\ast_{a}{{\cal T}^{\ast a}}_{bc}
+2{\cal D}^\ast_{[b}{\cal T}^\ast_{c]}
+{{\cal T}^{\ast e}}_{bc}{\cal T}^\ast_e
+2{{\cal R}}_{[bc]}
- 2{\cal H}_{bc}
& = & 0. \label{wgtcbi3}
\end{eqnarray}
\end{subequations}

\subsection{Manifestly covariant variational derivatives in WGT}

We now apply the manifestly covariant variational principle described
in Section~\ref{sec:mancovvp} to WGT. We begin by
deriving the variational derivatives,
and hence the EL equations, for the matter fields $\vpsi_A$ and
the gravitational gauge fields ${h_a}^\mu$,
${A^{ab}}_\mu$ and $B_\mu$. Using the fact that
$\delta_0h^{-1} = -h^{-1}{b^a}_\mu\,\delta_0{h_a}^\mu$, one may write
(\ref{eq:lagformvar2}) as
\bea
h\,\delta_0{\cal L}
& = & \delta_0L - {b^a}_\mu L\,\delta_0{h_a}^\mu,\nonumber \\
& = & 
\frac{\bar{\partial} L}{\partial\vpsi_A}\,\delta_0\vpsi_A + 
\pd{L}{({\cal D}^\ast_a\vpsi_A)}\,\delta_0({\cal D}^\ast_a\vpsi_A) +
\pd{L}{{\cal R}_{abcd}}\,\delta_0{\cal R}_{abcd} +
\pd{L}{{\cal T}^\ast_{abc}}\,\delta_0{\cal T}^\ast_{abc} +
\pd{L}{{\cal H}_{ab}}\,\delta_0{\cal H}_{ab}- {b^a}_\mu L\,\delta_0{h_a}^\mu.\phantom{AA}
\label{eq:hdL}
\eea
In order to progress further, one must determine how the variations in
(\ref{eq:hdL}) depend on the variations of the matter and
gravitational gauge fields themselves. This is easily achieved using
the definition of the WGT covariant derivative and the expressions
(\ref{rfsdef}--\ref{tfsdef}) for the field strengths. One must also 
make use of the fact that for any coordinate vector $V^\mu$ of
weight $w=0$ (i.e.\ invariant under local scale transformations, like
the Lagrangian density ${\cal L}$), one may show that $\partial_\mu
V^\mu = h^{-1}({\cal D}^\ast_a + {\cal T}^\ast_a)(h{b^a}_\mu V^\mu)$
or, equivalently, for any local Lorentz vector ${\cal V}^a$ having
Weyl weight $w = -3$ one has \cite{eWGTpaper}
\be
({\cal D}^\ast_a + {\cal T}^\ast_a){\cal V}^a = h\partial_\mu(h^{-1}
   {h_a}^\mu {\cal V}^a).
\label{eq:totald}
\ee
Such expressions on the RHS of (\ref{eq:hdL}) therefore contribute
only surface terms to the variation of the action in
(\ref{eq:genactioninv-mancov}), but we will retain them nonetheless, as
they are required for our later discussion.

We begin by considering together the first two terms on the RHS of
(\ref{eq:hdL}), for which one obtains
\bea
\frac{\bar{\partial} L}{\partial\vpsi_A}\,\delta_0\vpsi_A + 
\pd{L}{({\cal D}^\ast_a\vpsi_A)}\,\delta_0({\cal D}^\ast_a\vpsi_A)
& & \nonumber\\
& & \hspace*{-4.5cm} = \frac{\bar{\partial} L}{\partial\vpsi_A}\,\delta_0\vpsi_A +
\pd{L}{({\cal D}^\ast_a\vpsi_A)}\left[{\cal
    D}^\ast_a(\delta_0\vpsi_A)+\delta_0{h_a}^\mu D^\ast_\mu\vpsi_A
+{h_a}^\mu(w_A\,\delta_0B_\mu+\tfrac{1}{2}\delta_0{A^{bc}}_\mu\Sigma_{bc})\vpsi_A
\right],\nonumber\\
&& \hspace*{-4.5cm} = \left[\frac{\bar{\partial} L}{\partial\vpsi_A} 
- ({\cal D}^\ast_a 
+ {\cal T}^\ast_a)\pd{L}{({\cal D}^\ast_a\vpsi_A)}\right]\delta_0\vpsi_A
+ \pd{L}{({\cal D}^\ast_a\vpsi_A)}\left[\delta_0{h_a}^\mu D^\ast_\mu\vpsi_A
+{h_a}^\mu(w_A\,\delta_0B_\mu+\tfrac{1}{2}\delta_0{A^{bc}}_\mu\Sigma_{bc})\vpsi_A
\right]\phantom{AA}\nonumber\\
&&\qquad\quad\quad\,\, + \,({\cal D}^\ast_a + {\cal T}^\ast_a)\left[\pd{L}{({\cal D}^\ast_a\vpsi_A)}\delta_0\vpsi_A\right],
\label{eq:term1+2}
\eea
where the quantity in square brackets in the final term is readily
shown to have Weyl weight $w=-3$. Analysing the further terms
containing derivatives on the RHS of (\ref{eq:hdL}) in a similar
manner, one finds
\bea
\pd{L}{{\cal R}_{abcd}}\,\delta_0{\cal R}_{abcd}
& = & 2\pd{L}{{\cal R}_{abcd}}\left[R_{ab\mu d}\,\delta_0 {h_{c}}^\mu
+ {h_{d}}^\mu{\cal D}^\ast_{c}(\delta_0 A_{ab\mu})\right] \nonumber\\
& = &  2\pd{L}{{\cal R}_{abcd}}R_{ab[\mu d]}\,\delta_0{h_c}^\mu
+ \left[
2{h_c}^\mu({\cal D}^\ast_d+{\cal T}^\ast_d) + {h_e}^\mu {{\cal T}^{\ast e}}_{cd}
\right]\left(\pd{L}{{\cal R}_{abcd}}\right)\,\delta_0 A_{ab\mu}\nonumber\\
&& \hspace{3.4cm} -2\,({\cal D}^\ast_d+{\cal T}^\ast_d)\left[\pd{L}{{\cal R}_{abcd}}{h_c}^\mu\,\delta_0 A_{ab\mu}\right],
\label{eq:term3}\\[3mm]
\pd{L}{{\cal T}^\ast_{abc}}\,\delta_0{\cal T}^\ast_{abc} 
& = & 2\pd{L}{{\cal T}^\ast_{abc}}\left[ T^\ast_{a\mu\nu}{h_c}^\nu\,\delta_0{h_b}^\mu
+ {h_c}^\nu {\cal D}^\ast_b(\delta_0 b_{a\nu}) +
{h_b}^\mu\,\delta_0 A_{ac\mu} + \eta_{ac}{h_b}^\mu\,\delta_0 B_\mu\right]
\nonumber \\
& = & 2\pd{L}{{\cal T}^\ast_{abc}}
\left[(T^\ast_{a\mu\nu}{h_c}^\nu \delta_b^d-\tfrac{1}{2}{{\cal
      T}^{\ast d}}_{bc}b_{a\mu})\,\delta_0{h_d}^\mu + {h_b}^\mu\,\delta_0A_{ac\mu}
+\eta_{ac}{h_b}^\mu\,\delta_0 B_\mu\right]
\nonumber \\
&& 
- 2 ({\cal D}^\ast_c+{\cal T}^\ast_c)
\left(\pd{L}{{\cal T}^\ast_{abc}}\right)b_{a\mu}\,\delta_0
     {h_b}^\mu
+ 2({\cal D}^\ast_c+{\cal T}^\ast_c)
\left[\pd{L}{{\cal T}^\ast_{abc}}b_{a\mu}\,\delta_0
     {h_b}^\mu\right], \label{eq:term4} \\[3mm]
\pd{L}{{\cal H}_{ab}}\,\delta_0{\cal H}_{ab}
& = & 2\pd{L}{{\cal H}_{ab}}\left[H_{\mu\nu} {h_b}^\nu\,\delta_0 {h_a}^\mu
+{h_b}^\nu{\cal D}^\ast_a(\delta_0 B_\nu)\right] \nonumber \\
& = & 2\pd{L}{{\cal H}_{ab}}\left(H_{\mu\nu}{h_b}^\nu\,\delta_0 {h_a}^\mu
+\tfrac{1}{2}{{\cal T}^{\ast c}}_{ab}{h_c}^\nu\,\delta_0 B_\nu\right)
+2({\cal D}^\ast_b+{\cal T}^\ast_b)\left(\pd{L}{{\cal
    H}_{ab}}\right){h_a}^\nu\,\delta_0 B_\nu\nonumber \\
&& \hspace{3.4cm} -2({\cal D}^\ast_b+{\cal T}^\ast_b)\left[\pd{L}{{\cal
    H}_{ab}}{h_a}^\nu\,\delta_0 B_\nu\right].
\label{eq:term5}
\eea
In the above expressions it is assumed that the appropriate
antisymmetrisations, arising from the symmetries of the field strength
tensors, are performed when the RHS are evaluated. It is also easily
shown that the quantity in square brackets in each of the last terms
in (\ref{eq:term3}--\ref{eq:term5}) has Weyl weight $w=-3$, so
according to (\ref{eq:totald}) each
such term contributes a surface term to the variation of the action
(\ref{eq:genactioninv-mancov}).

One may then substitute the expressions
(\ref{eq:term1+2}--\ref{eq:term5}) into (\ref{eq:hdL}), which may
itself subsequently be substituted into (\ref{eq:genactioninv-mancov})
to obtain an expression of the general form
(\ref{eq:genactioninv-mancov2}) for Noether's first theorem, which
may be written as
\be
\delta S = \int \left[\upsilon^A\,\delta_0\vpsi_A
+ {\tau^a}_\mu\,\delta_0{h_a}^\mu 
+ {\sigma_{ab}}^\mu\,\delta_0{A^{ab}}_\mu
+ \zeta^\mu\,\delta_0B_\mu
+ h^{-1}({\cal D}^\ast_p + {\cal T}^\ast_p) (h{\cal J}^p)
\right]\,d^4x = 0,
\ee
where the current $h{\cal J}^p$ is given by
\be
h{\cal J}^p = \pd{L}{({\cal D}^\ast_p\vpsi_A)}\delta_0\vpsi_A
+2\left(\pd{L}{{\cal T}^\ast_{abp}}b_{a\mu}\,\delta_0 {h_b}^\mu
- \pd{L}{{\cal R}_{abcp}}{h_c}^\mu\,\delta_0 A_{ab\mu}
-\pd{L}{{\cal
    H}_{ap}}{h_a}^\mu\,\delta_0 B_\mu\right) + {b^p}_\mu \xi^\mu L,
\label{eq:wgtjcurrent}
\ee
and we have defined the variational derivative $\upsilon^A \equiv
\delta{\cal L}/\delta\vpsi_A$ with respect to the matter field
$\vpsi_A$, and the total dynamical energy-momentum ${\tau^a}_\mu
\equiv \delta{\cal L}/\delta {h_a}^\mu$, spin-angular-momentum
${\sigma_{ab}}^\mu \equiv \delta{\cal L}/\delta {A^{ab}}_\mu$ and
dilation current $\zeta^\mu \equiv \delta{\cal L}/\delta B_\mu$ of
both the matter and gravitational gauge fields.  Manifestly covariant
forms for these variational derivatives may be read off from the
expressions (\ref{eq:term1+2}--\ref{eq:term5}). Converting all Greek
indices to Roman and defining the quantities ${\tau^a}_b \equiv
{\tau^a}_\mu {h_b}^\mu$, ${\sigma_{ab}}^c \equiv {\sigma_{ab}}^\mu
{b^c}_\mu$ and $\zeta^a\equiv \zeta^\mu {b^a}_\mu$, one then makes the
following identifications
\begin{subequations}
\label{eq:mcvarderivs}
\bea
h\upsilon^A &=& \frac{\bar{\partial} L}{\partial\vpsi_A} 
- ({\cal D}^\ast_a + {\cal T}^\ast_a)\pd{L}{({\cal D}^\ast_a\vpsi_A)}, \label{eq:eomvpsi}\\[3mm]
h{\tau^a}_b & = & \pd{L}{({\cal D}^\ast_a\vpsi_A)}{\cal D}^\ast_b\vpsi_A
+ 2\pd{L}{{\cal R}_{pqra}}{\cal R}_{pqrb} 
+ 2\pd{L}{{\cal H}_{pa}}{\cal H}_{pb} 
+ 2\pd{L}{{\cal T}^\ast_{pqa}}{\cal T}^\ast_{pqb}
\!-\![{{\cal T}^{\ast a}}_{qr} + 2\delta_q^a
({\cal D}^\ast_r \!+\! {\cal T}^\ast_r)]\pd{L}{{{\cal T}^{\ast b}}_{qr}}
\!-\!\delta_a^b L,\phantom{AAa}\label{eq:eomh}\\[3mm]
h{\sigma_{ab}}^c 
& = &  \tfrac{1}{2}\pd{L}{({\cal D}^\ast_c\vpsi_A)}\Sigma_{ab}\vpsi_A
+\left[{{\cal T}^{\ast c}}_{pq} + 2\delta_p^c({\cal D}^\ast_q + {\cal T}^\ast_q)
\right]\pd{L}{{{\cal R}^{ab}}_{pq}}-2\pd{L}{{{\cal T}^{\ast [ab]}}_{c}},\label{eq:eoma}\\[3mm]
h\zeta^a & = & \pd{L}{({\cal D}^\ast_a\vpsi_A)}w_A\vpsi_A
 + \left[{{\cal T}^{\ast a}}_{pq} + 2 \delta_p^a({\cal D}^\ast_q + {\cal T}^\ast_q)
\right]\pd{L}{{\cal H}_{pq}}
+ 2\pd{L}{{{\cal T}^{\ast p}}_{qr}}\delta_q^a\delta_r^p,\label{eq:eomb}
\eea
\end{subequations}
where, once again, it is assumed that the appropriate
antisymmetrisations, arising from the symmetries of the field strength
tensors, are performed when the RHS are evaluated.  The expressions
(\ref{eq:mcvarderivs}) constitute the completion of our first
goal. One sees immediately that, unlike (\ref{eq:eleq}), the above
forms for the variational derivative of each field (and hence the
equations of motion obtained by setting each RHS to zero) are
manifestly covariant. Moreover, they are straightforward to evaluate,
since they require one only to differentiate the Lagrangian $L$ with
respect to the matter fields, their covariant derivatives and the
field strengths, respectively. One may easily confirm that the above expressions
lead to precisely the same variational derivatives as those obtained
by using the standard (but much longer) approach of evaluating
(\ref{eq:eleq}) for each field and then reassembling the many
resulting terms into manifestly covariant forms. 

The expressions (\ref{eq:mcvarderivs}) not only provide a significant
calculational saving in obtaining the variational derivatives in WGT,
but also yield a useful insight into their general form. In
particular, one notes that for a Lagrangian $L$ that does not contain
the gauge field strength tensors, but depends only on the matter
fields and their covariant derivatives, the variational derivatives
with respect to the gauge fields reduce to the {\it covariant
  canonical currents} \cite{Blagojevic02,CGTpaper} of the matter
fields. For Lagrangians that do depend on the gauge field strengths,
also of interest are the analogous forms of the penultimate terms on
the RHS of (\ref{eq:eomh}-\ref{eq:eomb}), which are the only terms
capable of producing a dependence on the covariant derivatives of the
field strength tensors; in each case, the corresponding term depends
on the covariant derivative of the field strength tensor for the gauge
field with respect to which the variational derivative is taken. It is
also worth pointing out that we have not assumed the equations of
motion to be satisfied in deriving
(\ref{eq:eomvpsi}--\ref{eq:eomb}). Thus, one may calculate the
corresponding variational derivatives for {\it any subset} of terms in
$L$ that is a scalar density of weight $w=-4$. Individually, however,
such quantities do {\it not} vanish, in general. Rather, each equation
of motion requires only the vanishing of the sum of such quantities,
when derived from disjoint subsets that exhaust the total Lagrangian
$L$.

Finally, we note that the above approach is easily adapted to other
gravitational gauge theories. For example, to apply it to PGT one
needs simply to `remove the asterisks', thereby replacing the WGT
covariant derivative and torsion by their PGT counterparts, and set
$B_\mu\equiv 0$, so that $\zeta^a$ and ${\cal H}_{ab}$ also vanish
identically. Indeed, the above approach is of even greater use in PGT
than WGT, since the functional dependence of the PGT Lagrangian on
the matter fields, their covariant derivatives and the field
strengths can be more complicated than in WGT, as in PGT one does not
require $L$ to have Weyl weight $w=-4$ \cite{Blagojevic02,eWGTpaper}.

\subsection{Relationship between first- and second-order variational
  derivatives in WGT}

Before turning our attention to the direct derivation of manifestly
covariant conservation laws for WGT, we first briefly demonstrate how
the analysis in the previous section is well suited to comparing
first- and second-order variational derivatives. In particular, we
will focus on the example of the variational derivatives obtained by
setting the WGT torsion to zero {\it after} the variation is performed
(first-order approach) with those obtained by setting the torsion to
zero in the action {\it before} carrying out the variation
(second-order approach).

Let us begin by considering the simpler case of the first-order
approach, where one merely sets ${{\cal T}^{\ast a}}_{bc} = 0$ (which
is a properly WGT-covariant condition) in the expressions
(\ref{eq:eomvpsi}--\ref{eq:eomb}). The condition ${{\cal T}^{\ast
    a}}_{bc} = 0$ results in the rotational gauge field ${A^{ab}}_\mu$
no longer being an independent field, but one determined explicitly by
the other gauge fields ${h_a}^\mu$ and $B_\mu$, which we thus denote
by $\zero{{A^{\ast ab}}_\mu}$ and term the `reduced' $A$-field
\cite{eWGTpaper,CGTpaper}. From (\ref{afromht}), these quantities
are given
explicitly by 
$\zero{A^\ast_{ab\mu}} = {b^c}_\mu \zero{{\cal A}^\ast_{abc}}$, where
\be
\zero{{\cal A}^\ast_{abc}} =
\tfrac{1}{2}(c_{abc}+c_{bca}-c_{cab})+\eta_{ac}{\cal B}_b -
\eta_{bc}{\cal B}_a,
\label{eq:afromht}
\ee
in which ${c^{a}}_{bc} \equiv {h_b}^\mu {h_c}^\nu (\partial_\mu
{b^a}_\nu-\partial_\nu {b^a}_\mu)$. Under a local Weyl transformation,
the quantities $\zero{{A^{\ast ab}}_\mu}$ transform in the same way as
${A^{ab}}_\mu$, so one may construct the `reduced' WGT covariant
derivative $\czero{{\cal D}}^\ast_a\vpsi = {h_a}^\mu
\,\zero{D^\ast_\mu} \vpsi = {h_a}^\mu (\partial_\mu +
\tfrac{1}{2}\zero{{A^{\ast cd}}_\mu}\Sigma_{cd}+wB_\mu)\vpsi$, which
transforms in the same way as ${\cal D}^\ast_a\vpsi$, but depends only
on the $h$ field, its first derivatives, and the $B$-field.  Thus, the
corresponding quantities to (\ref{eq:eomvpsi}--\ref{eq:eomb}) are
obtained simply by evaluating the RHS with ${{\cal T}^{\ast a}}_{bc}$
(and its contractions) set to zero, which also implies ${\cal
  D}^\ast_a \to \czero{{\cal D}}^\ast_a$. This yields
\begin{subequations}
\label{eq:eeom0}
\bea
h\,\zero{\upsilon}^A &=& 
\left.\frac{\bar{\partial} L}{\partial\vpsi_A}\right|_0
- \czero{{\cal D}^\ast_a} \left.\pd{L}{({\cal D}^\ast_a\vpsi_A)}\right|_0, \label{eq:eomvpsi0}\\[3mm]
h\,\zero{{\tau^a}_b} & = & 
\left.\pd{L}{({\cal D}^\ast_a\vpsi_A)}\right|_0
\czero{{\cal D}^\ast_b}\vpsi_A
+ 2\left.\pd{L}{{\cal R}_{pqra}}\right|_0
\czero{{\cal R}}_{pqrb} 
+ 2\left.\pd{L}{{\cal H}_{pa}}\right|_0{\cal
  H}_{pb}
+ 2\,\czero{{\cal D}^\ast_r}\left.\pd{L}{{{\cal T}^{\ast b}}_{ar}}\right|_0 
-\delta_a^b \left.L\right|_0,\label{eq:eomh0}\\[3mm]
h\,\zero{{\sigma_{ab}}^c} 
& = &  \tfrac{1}{2}\left.\pd{L}{({\cal D}^\ast_c\vpsi_A)}\right|_0\Sigma_{ab}\vpsi_A
+2\delta_r^c\,\czero{{\cal D}^\ast_s}
\left.\pd{L}{{{\cal R}^{ab}}_{rs}}\right|_0
-\left.2\pd{L}{{{\cal T}^{\ast [ab]}}_{c}}\right|_0,\label{eq:eoma0}\\[3mm]
h\,\zero{\zeta^a} & = & \left.\pd{L}{({\cal D}^\ast_a\vpsi_A)}\right|_0 w_A\vpsi_A
+ 2 \delta_p^a\,\czero{{\cal D}^\ast_q}
\left.\pd{L}{{\cal H}_{pq}}\right|_0+ 2\left.\pd{L}{{{\cal T}^{\ast p}}_{qr}}\right|_0 \delta_q^a\delta_r^p 
,\label{eq:eomb0}
\eea
\end{subequations}
where $|_0$ denotes that the quantity to its immediate left is
evaluated assuming ${\cal T}^\ast_{abc}=0$.  The equations of motion from the
first-order approach are then given simply by equating each of
(\ref{eq:eomvpsi0}--\ref{eq:eomb0}) to zero. 
Once again, it is worth noting that we have not assumed any equations
of motion to be satisfied in deriving the quantities
(\ref{eq:eomvpsi0}-- \ref{eq:eomb0}). Thus, one may derive
corresponding quantities for {\it any subset} of terms in $L$ that are
a scalar density with weight $w=-4$, and these quantities do not
vanish, in general.

We now consider the second-order approach, where one imposes ${\cal
    T}^{\ast}_{abc} = 0$ at the level of action, prior to evaluating
the variational derivatives. In this case, the rotational gauge field
${A^{ab}}_\mu$ is again determined explicitly by ${h_a}^\mu$ and
$B_\mu$ according to (\ref{eq:afromht}), and so now the action depends
only on these other gauge fields. From (\ref{eq:afromht}), one readily
finds that
\be
\delta_0 A_{ab\mu} = {b^c}_\mu \left(
{h_{[c}}^\nu\,\czero{{\cal D}^\ast_{b]}}\delta_0b_{a\nu}
+{h_{[a}}^\nu\,\czero{{\cal D}^\ast_{c]}}\delta_0 b_{b\nu}
-{h_{[b}}^\nu\,\czero{{\cal D}^\ast_{a]}}\delta_0 b_{c\nu}
+2\eta_{c[a} {h_{b]}}^\nu \delta_0B_\nu\right),
\ee
from which one may show that (up to terms that are the divergence of a
quantity that vanishes on the boundary of the integration region) the
integrand in the expression (\ref{eq:genactioninv}) for the variation
of the action is given by
\bea
\frac{\delta{\cal L}}{\delta\chi_A}\,\delta_0\chi_A
&=& \zero{\upsilon}^A\,\delta_0\vpsi_A
+ \czero{{\tilde{\tau}^a}_{\phantom{a}\mu}}\,\delta_0{h_a}^\mu 
-b{b^f}_\mu\left(
\eta_{fa}\delta_{[b}^e\czero{{\cal D}^\ast_{c]}}
+\eta_{fb}\delta_{[c}^e\czero{{\cal D}^\ast_{a]}}
-\eta_{fc}\delta_{[a}^e\czero{{\cal D}^\ast_{b]}}\right)
(h\,\czero{\tilde{\sigma}}^{abc})\,\delta_0{h_e}^\mu\nonumber \\
&&\hspace{7.6cm} +
2\,\czero{\tilde{\sigma}}^{abc}\eta_{c[a}{h_{b]}}^\nu\,\delta_0 B_\mu
+ \czero{\tilde{\zeta}}^\mu\,\delta_0 B_\mu,\label{eq:master0}\\
& \equiv & v^A\,\delta_0\vpsi_A
+ {t^a}_\mu\,\delta_0 {h_a}^\mu 
+ j^\mu\,\delta_0 B_\mu,\label{eq:master1}
\eea
where we have again made use of (\ref{eq:totald}) and
$\czero{{\tilde{\tau}^a}_{\phantom{a}\mu}}$,
$\czero{\tilde{\sigma}_{ab}^{\phantom{ab}c}}$ and
$\czero{\tilde{\zeta}}^\mu$ denote quantities analogous to
(\ref{eq:eomh0}--\ref{eq:eomb0}), respectively, but {\it without} the
terms containing $\left.\partial L/\partial {\cal T}^\ast_{abc}\right|_0$. In the
last line, we have also defined the total dynamical energy-momentum
${t^a}_\mu$ and dilation current $j^\mu$ of both the matter and
gravitational gauge fields, and the matter field variational
derivatives $v^A$, in the second-order approach. By comparing
(\ref{eq:master0}) and (\ref{eq:master1}), and converting all indices
to Roman, one finds that the second-order variational derivatives are
given in terms of the first-order ones by
\bea
hv^A & = & h\,\zero{\upsilon}^A,\label{eq:eomspsi}\\
ht_{ab} & = & h\,\czero{{\tilde{\tau}}_{ab}} + \czero{{\cal D}}^\ast_c
\left(h\,\czero{\tilde{\sigma}}_{ab}^{\phantom{ab}c}
-h\,\czero{\tilde{\sigma}}^c_{\phantom{c}ab}
-h\,\czero{\tilde{\sigma}}^c_{\phantom{c}ba}\right),\label{eq:eomsh}\\
hj^a & = & h(\czero{\tilde{\zeta}}^a - 2\,\czero{\tilde{\sigma}}^{ab}_{\phantom{ab}b}).
\label{eq:eomsb}
\eea

It is clear that the forms of the matter variational derivatives are
identical in the two approaches, but those of the gravitational gauge
fields ${h_a}^\mu$ and $B_\mu$ differ, in general. In particular, the
form for the energy-momentum tensor $t_{ab}$ in the second-order
approach is clearly the gauge theory equivalent of the Belinfante
tensor \cite{Belinfante40}.
By analogy, the expression
(\ref{eq:eomsb}) may be considered to define an associated Belinfante
dilation current, which is clearly related to the `field virial' that
is relevant to the invariance of an action under special conformal
transformations \cite{Coleman71,CGTpaper}.

It is again worth noting that the expressions
(\ref{eq:eomspsi}--\ref{eq:eomsb}) have been derived without assuming
any equations of motion are satisfied. One may therefore obtain
analogous relations between corresponding first- and second-order
variational derivatives derived from {\it any subset} of the terms in
the total Lagrangian $L$ that are a scalar density of weight $w=-4$.
If one does consider the total Lagrangian $L$, however, then the
second-order equations of motion for the matter and gauge fields are
obtained simply by setting the expressions
(\ref{eq:eomspsi}--\ref{eq:eomsb}) to zero. In this case, provided the
terms of the form $\left.\partial L/\partial {\cal
  T}^\ast_{abc}\right|_0$ vanish in the first-order equations of
motion obtained by setting (\ref{eq:eeom0}--\ref{eq:eomb0}) to zero,
then this implies that the second-order equations of motion obtained
by setting (\ref{eq:eomspsi}-\ref{eq:eomsb}) to zero are also
satisfied, but the contrary does not necessarily hold.

\subsection{Manifestly covariant conservation laws in WGT}
\label{sec:mcconswgt}

We now turn our attention to deriving the conservation laws for WGT in
a manner that maintains manifest covariance throughout, by applying
the general method outlined in Section~\ref{sec:mancovvp}. One may
begin by considering the general form of the conservations laws given
in (\ref{eq:n2cond1-mancov}). As discussed above, the key issue to
address is the forms of the functions $f_{AC}$ and $f^\mu_{AC}$ that
appear in this expression and define the form variations
(\ref{eq:localformvar}) of the fields, since these are typically not
covariant. For (\ref{eq:n2cond1-mancov}) to be valid, one requires at
least the functions $f^\mu_{AC}$ to be (manifestly) covariant,
although many of the functions $f_{AC}$ may also be made so; as
outlined in Section~\ref{sec:mancovvp}, this is
performed by generalising the approach introduced by
Bessel-Hagen for electromagnetism, which is reviewed in
Appendix~\ref{app:BHmethod}, and developed further below.

The form variations of the fields in WGT are given in
(\ref{eq:wgtfieldtrans}). By comparing these transformation laws with
the generic form (\ref{eq:localformvar}), one may read off the
functions $f_{AC}$ and $f_{AC}^\mu$ in the latter from the
coefficients of $\{\lambda^C\} = \{\lambda^1,\lambda^2,\lambda^3\} = \{\xi^\alpha,
\omega^{ab},\rho\}$ and their partial derivatives, respectively. As
anticipated, one immediately finds that many of the functions $f_{AC}$
and $f_{AC}^\mu$ are not covariant quantities.  In the context of the
Bessel-Hagen method, the form variations (\ref{eq:wgtfieldtrans}) are
already in the most general form that leaves the generic WGT action
invariant (ignoring the possibility of additional accidental
symmetries occurring). Following the general methodology outlined for
electromagnetism in Appendix~\ref{app:BHmethod}, we consider
separately the conservation laws that result from the invariance of
the WGT action under infinitesimal GCTs, local rotations and local
dilations, respectively.

Considering first the infinitesimal GCTs characterised by
$\xi^\alpha(x)$ (which we take to correspond to $C=1$), one may make use
of the invariance of the action under the transformations
(\ref{eq:wgtfieldtrans}) for arbitrary functions $\omega^{ab}(x)$ and
$\rho(x)$ by choosing them in a way that yields covariant forms for
the new functions $f_{A1}^\mu$ (and also $f_{A1}$ in this case) in the
resulting form variations. This is achieved by setting 
$\omega^{ab} = -{A^{ab}}_\nu\xi^\nu$ and $\rho = -B_\nu\xi^\nu$ (where
the minus signs are included for later convenience), which yields 
transformation laws of a much simpler form than in 
(\ref{eq:wgtfieldtrans}), given by
\begin{subequations}
\label{eq:wgtfieldtransnew}
\bea
\delta_0\vpsi & = & -\xi^\nu D^\ast_\nu\vpsi,
\label{eq:vpsitransnew}\\
\delta_0{h_a}^\mu & = & 
-\xi^\nu D^\ast_\nu {h_a}^\mu + {h_a}^\nu\partial_\nu\xi^\mu,
\label{eq:htransnew}\\
\delta_0 {A^{ab}}_\mu & = & \xi^\nu {R^{ab}}_{\mu\nu}, \label{eq:atransnew}\\
\delta_0 B_\mu & = &  \xi^\nu H_{\mu\nu}.\label{eq:btransnew}
\eea
\end{subequations}
From these form variations, one may immediately read off the new forms
of the functions $f_{A1}$ and $f^\mu_{A1}$, all of which are now
manifestly covariant. Inserting these expressions into the general
form (\ref{eq:n2cond1-mancov}), one directly obtains the manifestly
covariant conservation law
\be
({\cal D}^\ast_c+{\cal T}^\ast_c)(h{\tau^c}_\nu) 
- h({\sigma_{ab}}^\mu {R^{ab}}_{\mu\nu} + \zeta^\mu H_{\mu\nu} - 
{\tau^a}_\mu D^\ast_\nu h_a^\mu - \upsilon^A D^\ast_\nu\vpsi_A) = 0,
\ee
where it is worth noting that $h\upsilon^A = \delta L/\delta\vpsi_A$.
On multiplying through by ${h_d}^\nu$, one may rewrite the
conservation law wholly in term of quantities possessing only Roman
indices as
\be
({\cal D}^\ast_c+{\cal T}^\ast_c)(h{\tau^c}_d) 
- h({\sigma_{ab}}^c {{\cal R}^{ab}}_{cd} + \zeta^c {\cal H}_{cd} 
- {\tau^c}_b {{\cal T}^{\ast\,b}}_{cd} - \upsilon^A {\cal D}^\ast_d\vpsi_A) = 0.
\label{eq:wgtcons1}
\ee

We next consider invariance of the action under infinitesimal local
Lorentz rotations characterised by $\omega^{ab}(x)$ (which we take to
correspond to $C=2$). In this case, the functions $f^\mu_{A2}$ in the
original set of transformation laws (\ref{eq:wgtfieldtrans}) are
already manifestly covariant. One may thus insert the functions
$f^\mu_{A2}$ and $f_{A2}$ read off from (\ref{eq:wgtfieldtrans})
directly into the general form (\ref{eq:n2cond1-mancov}), without
employing the Bessel-Hagen method. On recalling that $\Gamma^\ast_\beta
{\sigma_{pq}}^\beta = -{A^r}_{p\beta}{\sigma_{rq}}^\beta -
{A^r}_{q\beta}{\sigma_{pr}}^\beta$ (since ${\sigma_{ab}}^\mu$ has Weyl
weight $w=0$) one finds that the final set of terms on the LHS of
(\ref{eq:n2cond1-mancov}) vanish when $\gamma^A$ corresponds to $
h{\sigma_{ab}}^\mu$, and one immediately obtains
the manifestly covariant conservation law
\be 
({\cal D}^\ast_c+{\cal T}^\ast_c)(h{\sigma_{ab}}^c) + h\tau_{[ab]}
+ \tfrac{1}{2}h\upsilon^A\Sigma_{ab}\vpsi_A = 0.
\label{eq:wgtcons2}
\ee

Finally, we consider invariance of the action under infinitesimal
local dilations characterised by $\rho(x)$ (which we take to
correspond to $C=3$). Once again, the relevant functions $f^\mu_{A3}$ in the
original set of transformation laws (\ref{eq:wgtfieldtrans}) are
already manifestly covariant. One may thus insert
$f^\mu_{A3}$ and $f_{A3}$ read off from (\ref{eq:wgtfieldtrans})
directly into the general form (\ref{eq:n2cond1-mancov}), which
immediately yields the manifestly covariant conservation law
\be 
({\cal D}^\ast_c+{\cal T}^\ast_c)(h\zeta^c) - h{\tau^c}_c
+ h\upsilon^A w_A\vpsi_A = 0.
\label{eq:wgtcons3}
\ee

It is straightforward to verify that the manifestly covariant
conservations WGT laws (\ref{eq:wgtcons1}--\ref{eq:wgtcons2}) have the
correct forms \cite{eWGTpaper,CGTpaper} and match those derived
(albeit at considerably greater length) using the standard form of
Noether's second theorem (\ref{eq:n2cond1}).
Before moving on to consider the further condition (\ref{eq:n2cond2}) arising
from Noether's second theorem, in the context of WGT, we note that the
conservation law (\ref{eq:wgtcons2}) may be used to simplify the
expression (\ref{eq:eomsh}) for the second-order variational
derivative with respect to ${h_a}^\mu$ in terms of first-order
variational derivatives. Imposing the condition ${\cal T}^\ast_{abc} =
0$, the conservation law (\ref{eq:wgtcons2}) becomes
\be
\czero{{\cal D}}^\ast_c(h\,\czero{\tilde{\sigma}_{ab}}^{\phantom{ab}c})
+ h\,\czero{\tilde{\tau}}_{[ab]} +\tfrac{1}{2}h\,\czero{\tilde{\upsilon}}^A
\Sigma_{ab}\vpsi_A  =  0.
\ee
If one assumes the {\it matter} equations of motion
$\czero{\tilde{\upsilon}}^A=0$ are satisfied (or, equivalently, that
the Lagrangian $L$ does not depend on matter fields), the expression
(\ref{eq:eomsh}) can thus be written in the simpler and 
manifestly symmetric form
\be
ht_{ab}  \eomm  h\,\czero{{\tilde{\tau}}_{(ab)}} -2\,\czero{{\cal
    D}}^\ast_c(h\,\czero{\tilde{\sigma}}^c_{\phantom{c}(ab)}).
\ee
%

\subsection{Relationship between currents in Noether's second theorem
  in WGT}

We conclude this section by considering the relationship in WGT
between the two currents that appear in Noether's second theorem
(\ref{eq:n2cond2}). As discussed in Section~\ref{subsec:wgtnt2}, this
equation may be re-written as $({\cal D}^\ast_a + {\cal T}^\ast_a)
[h({\cal J}^a -{\cal S}^a)] = 0$, where $h{\cal J}^a$ for WGT is given
by (\ref{eq:wgtjcurrent}) and the expression for $h{\cal S}^a$ may be
obtained from the general form (\ref{eq:generalscurrent}), which on
using the original WGT field variations (\ref{eq:wgtfieldtrans})
yields
\be
h{\cal S}^p = h\left[-\xi^\mu({\tau^p}_\mu - {\sigma_{ab}}^p
  {A^{ab}}_\mu - \zeta^p B_\mu) + \omega^{ab}{\sigma_{ab}}^p +
  \rho \zeta^p\right]. 
\label{eq:wgtscurrent}
\ee
It is worth noting that this expression does not depend on the variational
derivatives $\upsilon^A \equiv \delta{\cal L}/\delta\psi_A$ with
respect to the matter fields since, as expected, the functions
$f^\mu_{AC}$ vanish in this case, as can be read off from the field
variations (\ref{eq:wgtfieldtrans}). Thus, in order for $h{\cal S}^p$
to vanish, it is sufficient that just the equations of motion of the gauge
fields are satisfied.

If one substitutes the original form variations
(\ref{eq:wgtfieldtrans}) into the expression (\ref{eq:wgtjcurrent})
for $h{\cal J}^p$, one finds after a long calculation\footnote{The
  calculation can be somewhat shortened, better organised and carried
  out in a largely manifestly covariant manner if one assumes the local
  Weyl transformaton parameters in (\ref{eq:wgtfieldtrans}) to have
  the forms $\xi^\mu(x)$, $\omega^{ab}(x) = \bar{\omega}^{ab}(x) -
  {A^{ab}}_\nu\xi^\nu$ and $\rho(x) = \bar{\rho}(x)-B_\nu\xi^\nu$,
  where $\xi^\mu(x)$, $\bar{\omega}^{ab}(x)$ and $\bar{\rho}(x)$ are
  arbitrary functions of position, and considers separately
  the three cases: (i) $\bar{\omega}^{ab} = 0 = \bar{\rho}$; (ii) 
$\xi^\mu = 0 = \bar{\rho}$; and (iii) $\xi^\mu = 0
  = \bar{\omega}^{ab}$.  This is a similar approach to that used in
  Section~\ref{sec:mcconswgt} to derive directly the manifestly
  covariant forms of the WGT conservation laws and, in particular,
  allows one in case (i) to make use again of the manifestly covariant
  form variations (\ref{eq:wgtfieldtransnew}) derived using the
  Bessel-Hagen method.}, which requires careful use of the definition
(\ref{eq:dcommwgt}) of the field strength tensors, the contracted
Bianchi identity (\ref{wgtcbi3}) and the manifestly covariant
expressions (\ref{eq:eomh}--\ref{eq:eomb}) for the variational
derivatives with respect to the gravitational gauge fields, that
\be
({\cal D}^\ast_p + {\cal T}^\ast_p)
(h{\cal J}^p) = 
({\cal D}^\ast_p + {\cal T}^\ast_p)
\left[-\xi^\mu h({\tau^p}_q{b^q}_\mu - {\sigma_{ab}}^p
  {A^{ab}}_\mu - \zeta^p B_\mu) + \omega^{ab}h{\sigma_{ab}}^p +
  \rho h\zeta^p\right]
= ({\cal D}^\ast_p + {\cal T}^\ast_p)
(h{\cal S}^p),
\ee
thereby verifying explicitly the relationship between the two currents
that is implied by Noether's second theorem (\ref{eq:n2cond2}). Thus,
as expected for an action that is invariant under a set of local
symmetries, this relationship contains no further information, but
nonetheless provides a useful check of the derivation of the
expressions (\ref{eq:eomh}--\ref{eq:eomb}). Indeed, the requirement
$({\cal D}^\ast_a + {\cal T}^\ast_a) [h({\cal J}^a -{\cal S}^a)] = 0$
from Noether's second theorem can thus be used as an alternative
(albeit rather longer) means of deriving the expressions
(\ref{eq:eomh}--\ref{eq:eomb}) for the variational derivatives with
respect to the gravitational gauge fields; it has been demonstrated,
however, that this equivalence between the Noether and Hilbert
(variational) approaches does not hold in general for all modified
gravity theories \cite{Baker21}.

\section{Extended Weyl gauge theory}
\label{sec:ewgt}

We now move on to consider eWGT \cite{eWGTpaper}, which proposes an
`extended' form for the transformation laws of the rotational and
dilational gauge fields under local dilations. In particular, under
infinitesimal local Weyl transformations consisting of GCTs, rotations
of the local Lorentz frames and dilations, parameterised by
$\xi^\mu(x)$, $\omega^{ab}(x)$ and $\rho(x)$, respectively, a matter
field $\vpsi$ of weight $w$ and the gauge fields transform as
\begin{subequations}
\label{eq:ewgtfieldtrans}
\bea
\delta_0\vpsi & = & -\xi^\nu\partial_\nu\vpsi +
(\tfrac{1}{2}\omega^{ab}\Sigma_{ab} + w\rho)\vpsi,
\label{eq:evpsitrans}\\
\delta_0{h_a}^\mu & = & 
-\xi^\nu\partial_\nu {h_a}^\mu + {h_a}^\nu\partial_\nu\xi^\mu
- ({\omega^b}_{a} +\rho\,\delta^b_a){h_b}^\mu,\phantom{AAA} 
\label{eq:ehtrans}\\
\delta_0 {A^{ab}}_\mu & = &-\xi^\nu\partial_\nu{A^{ab}}_\mu
-{A^{ab}}_\nu \partial_\mu\xi^\nu
-2{{\omega^{[a}}_{c}A^{b]c}}_\mu 
-\partial_\mu\omega^{ab} + 2\theta\eta^{c[a}{b^{b]}}_\mu {h_c}^\nu\partial_\nu\rho, \label{eq:eatrans}\\
\delta_0 B_\mu & = &  -\xi^\nu\partial_\nu B_\mu -B_\nu\partial_\mu\xi^\nu
-\theta\partial_\mu\rho,\label{eq:ebtrans}
\eea
\end{subequations}
where $\theta$ is an arbitrary parameter that can take any value.  The
proposed form for the transformation law (\ref{eq:eatrans}) of the
rotational gauge field is motivated by the observation that the WGT
(and PGT) matter actions for the massless Dirac field and the
electromagnetic field (neither of which depends on the dilation gauge
field) are invariant under local dilations even if one assumes this
`extended' transformation law for the rotational gauge field. A
complementary motivation for introducing the extended transformation
law (\ref{eq:eatrans}) is that under local dilations it places the
transformation properties of the PGT rotational gauge field strength
${{\cal R}^{ab}}_{cd}$ and translational gauge field strength ${{\cal
    T}^a}_{bc}$ on a more equal footing: for general values of
$\theta$, neither ${{\cal R}^{ab}}_{cd}$ nor ${{\cal T}^{a}}_{bc}$
transforms covariantly, but ${{\cal R}^{ab}}_{cd}$ does transform
covariantly and ${{\cal T}^a}_{bc}$ transforms inhomogeneously for
$\theta=0$, and {\it vice-versa} for $\theta = 1$. It is also worth
noting that the extended transformation law for the rotational gauge field
reduces to that in WGT for $\theta=0$, whereas the extended transformation law
(\ref{eq:ebtrans}) for the dilational gauge field reduces to the WGT
form for $\theta=1$; thus there is no value of $\theta$ for which both
transformation laws reduce to their WGT forms.

In eWGT, the covariant derivative, denoted by ${\cal
  D}^\dagger_a$, has a somewhat different form to that shown in
(\ref{eq:wgtcovderivdef}) for WGT. In particular, one does not adopt
the standard approach of introducing each gauge field as the linear
coefficient of the corresponding generator. Rather, in order to
accommodate our proposed extended transformation law
(\ref{eq:eatrans}) under local dilations, one is led to introduce the
`rotational' gauge field ${A^{ab}}_\mu(x)$ and the `dilational' gauge
field $B_\mu(x)$ in a very different way, so that\footnote{The
  daggers in the definition of the derivative operator are intended
  simply to distinguish it from the usual notation used
  \cite{Blagojevic02,eWGTpaper,CGTpaper} for the covariant derivatives
  of PGT and WGT, and should not be confused with the operation of
  Hermitian conjugation.}
\be
{\cal D}^\dagger_a\vpsi_A = {h_a}^\mu D^\dagger_\mu \vpsi_A
= {h_a}^\mu (\partial_\mu + \Gamma^\dagger_\mu)\vpsi_A
= {h_a}^\mu [\partial_\mu +
\tfrac{1}{2}{A^{\dagger ab}}_\mu\Sigma_{ab}+w_A(B_\mu-\tfrac{1}{3}T_\mu)]\vpsi_A,
\label{eq:ewgtgencovdef}
\ee
where  we have introduced
the modified $A$-field
\be
{A^{\dagger ab}}_\mu  \equiv  {A^{ab}}_\mu + 2{b^{[a}}_\mu{\cal B}^{b]},
\label{adaggerdef}
\ee
in which ${\cal B}_a = {h_a}^\mu B_\mu$ and $T_\mu = {b^a}_\mu {\cal
  T}_a$, where ${\cal T}_a \equiv {{\cal T}^b}_{ab}$ is the trace of
the PGT torsion.\footnote{It is worth noting that ${A^{\dagger
    ab}}_\mu$ is not considered to be a fundamental field
(notwithstanding the variational approach adopted below), but merely a
shorthand for the above combination of the gauge fields ${h_a}^\mu$
(or its inverse), ${A^{ab}}_\mu$ and $B_\mu$. Similarly, $T_\mu$ is
merely a shorthand for the corresponding function of the gauge fields
${h_a}^\mu$ (or its inverse) and ${A^{ab}}_\mu$.}  It is
straightforward to show that, if $\vpsi$ has Weyl weight $w$, then
(\ref{eq:ewgtgencovdef}) does indeed transform covariantly with Weyl
weight $w-1$, as required.  Unlike the transformation laws for
${A^{ab}}_\mu$ and $B_\mu$, the covariant derivative
(\ref{eq:ewgtgencovdef}) does not explicitly contain the parameter
$\theta$. Consequently, it does {\it not} reduce to the standard WGT
covariant derivative ${\cal D}_a^\ast\vpsi_A$ in either special case
$\theta=0$ or $\theta=1$, while retaining its covariant transformation
law for {\it any} value of $\theta$.

The derivative (\ref{eq:ewgtgencovdef}) does in fact transform
covariantly under the much {\it wider} class of gauge field
transformations in which $\theta \partial_\mu\rho(x)$ is replaced in
(\ref{eq:eatrans}--\ref{eq:ebtrans}) by an {\it arbitrary} vector
field $Y_\mu(x)$. Indeed, one finds that the WGT (and PGT) matter
actions for the massless Dirac field and the electromagnetic field are
still invariant under local dilations after such a replacement,
although the discussion above regarding the transformation properties
of ${{\cal R}^{ab}}_{cd}$ and ${{\cal T}^a}_{bc}$ following requires
appropriate modification, since neither transforms covariantly if
$\theta\partial_\mu \rho(x)$ is replaced by an arbitrary vector
$Y_\mu(x)$. The covariance of ${\cal D}^\dagger_a\vpsi_A$ under this
wider class of transformations allows one to identify a further gauge
symmetry of eWGT, namely under the simultaneous transformations
\be
   {A^{ab}}_\mu \to {A^{ab}}_\mu + 2{b^{[a}}_\mu{\cal Y}^{b]}, \qquad
B_\mu \to B_\mu - Y_\mu,
\label{eq:tsg}
\ee
where ${\cal Y}_a = {h_a}^\mu
Y_\mu$ and $Y_\mu$ is an arbitrary vector field. Under this symmetry,
both ${A^{\dagger ab}}_\mu$ and $B_\mu - \tfrac{1}{3}T_\mu$ remain
unchanged and thus ${\cal D}^\dagger_a\vpsi$ is invariant, as too are
the eWGT field strengths and action discussed below. One may make use
of this symmetry of eWGT to choose a gauge in which either $B_\mu$ or
$T_\mu$ is self-consistently set to zero, which can considerably
simplify subsequent calculations.

It was noted in \cite{eWGTpaper} that the extended transformation
laws (\ref{eq:eatrans}--\ref{eq:ebtrans}) implement Weyl scaling in a
novel way that may be related to gauging of the full conformal group.
This is discussed in more detail in \cite{CGTpaper}, where it is shown
that eWGT does indeed constitute a valid novel gauge theory of the
conformal group. We briefly summarise below the aspects of eWGT
that are relevant to our present discussion, and refer the reader to
\cite{eWGTpaper,CGTpaper} for further details.

By analogy with WGT, the Lagrangian density in eWGT has the usual form
${\cal L} = h^{-1}L$, where the translational gauge field ${h_a}^\mu$
is assigned a Weyl weight $w=-1$, so that $h = {\rm det}({h_a}^\mu)$
and $L$ are scalar densities both of Weyl weight $w=-4$, and hence the
action $S$ is invariant under local scale transformations. The Lagrangian has
the functional dependencies
\be
L = L(\vpsi_A, {\cal D}^\dagger_a\vpsi_A,{\cal R}^\dagger_{abcd},{\cal
  T}^\dagger_{abc},{\cal H}^\dagger_{ab}),
\label{eq:ewgtlag}
\ee
where the quantities ${\cal R}^\dagger_{abcd}$, ${\cal
  T}^\dagger_{abc}$, ${\cal H}^\dagger_{ab}$ are the eWGT
`rotational', `translational' and `dilational' gauge field strengths,
respectively, which are defined through the action of the commutator
of two eWGT covariant derivatives on some field $\vpsi$ of weight $w$ by
\be
[{\cal D}^\dagger_c,{\cal D}^\dagger_d]\vpsi = (\tfrac{1}{2}{{\cal R}^{\dagger
ab}}_{cd}\Sigma_{ab} + w{\cal H}^\dagger_{cd}
- {{\cal T}^{\dagger a}}_{cd}{\cal D}^\dagger_a) \vpsi.
\label{eq:dcommewgt}
\ee
The field strengths have the forms ${{\cal R}^{\dagger ab}}_{cd}
= {h_a}^{\mu}{h_b}^{\nu}{R^{\dagger ab}}_{\mu\nu}$, ${\cal H}^\dagger_{cd} =
       {h_c}^\mu {h_d}^\nu H^\dagger_{\mu\nu}$ and ${{\cal T}^{\dagger a}}_{bc} =  {h_b}^{\mu}{h_c}^{\nu} {T^{\dagger a}}_{\mu\nu}$, where
\bea
{R^{\dagger ab}}_{\mu\nu}  & = & 2(\partial_{[\mu} {A^{\dagger ab}}_{\nu]}
+\eta_{cd}{A^{\dagger ac}}_{[\mu}{A^{\dagger db}}_{\nu]}),
\label{erfsdef} \\
H^\dagger_{\mu\nu} & = & 2(\partial_{[\mu} B_{\nu]}-\tfrac{1}{3}\partial_{[\mu} T_{\nu]}),
\label{edfsdef}\\
{T^{\dagger a}}_{\mu\nu} & = & 2D^\dagger_{[\mu} {b^a}_{\nu]}.
\label{etfsdef}
\eea
From the transformation laws (\ref{eq:ewgtfieldtrans}), it is
straightforward to verify that, in accordance with their index
structures, the gauge field strength tensors ${{\cal R}^{\dagger ab}}_{cd}$,
${\cal H}^\dagger_{cd}$ and ${{\cal T}^{\dagger a}}_{bc}$ are invariant under
GCTs, and transform covariantly under local Lorentz transformations
and dilations with Weyl weights $w = -2$, $w=-2$ and $w=-1$,
respectively \cite{eWGTpaper,CGTpaper}, similarly to their WGT counterparts.

It is worth noting, however, that 
${{\cal R}^{\dagger ab}}_{cd}$ and ${{\cal T}^{\dagger a}}_{bc}$
differ in form substantially from those in WGT, and
are given in terms of the PGT field strengths ${{\cal R}^{ab}}_{cd}$
and ${{\cal T}^{a}}_{bc}$ by
\bea
{{\cal R}^{\dagger ab}}_{cd} & \!\!=\! & \!{{\cal R}^{ab}}_{cd}
\!+\!4\delta^{[b}_{[c}  {\cal D}_{d]}  {\cal B}^{a]} \!-\!
4\delta^{[b}_{[c} {\cal B}_{d]}{\cal B}^{a]}
\!-\!2 {\cal B}^2 \delta^{[a}_c \delta^{b]}_d
\!-\!2 {\cal B}^{[a}{{\cal T}^{b]}}_{cd},\nonumber \\
{{\cal T}^{\dagger a}}_{bc}  & = &  {{\cal T}^a}_{bc}
+\tfrac{2}{3}\delta^a_{[b} {\cal T}_{c]},
\eea
where ${\cal B}^2 \equiv {\cal B}^a {\cal B}_a$ and ${\cal D}_a \equiv
{h_a}^\mu D_\mu \equiv {h_a}^\mu(\partial_\mu +
\tfrac{1}{2}{A^{ab}}_\mu\Sigma_{ab})$ is the PGT covariant derivative
operator.  It is particularly important to note that the trace of the
eWGT translational field strength tensor vanishes identically, namely
${\cal T}^\dagger_b \equiv {{\cal T}^{\dagger a}}_{ba} = 0$, so that
${{\cal T}^{\dagger a}}_{bc}$ is completely trace-free (contraction on
any pair of indices yields zero).
Moreover, using the expression (\ref{etfsdef}) and defining the
quantities ${c^{\ast a}}_{bc} \equiv 2{h_b}^\mu {h_c}^\nu
\partial^\dagger_{[\mu} {b^a}_{\nu]}$, where $\partial^\dagger_\mu =
\partial_\mu + w (B_\mu-\tfrac{1}{3}T_\mu)$, one may show that the fully
anholonomic modified $A$-field ${{\cal A}^{\dagger ab}}_c \equiv {h_c}^\mu
{A^{\dagger ab}}_\mu$ can be written as \cite{eWGTpaper}
\be
{\cal A}^\dagger_{abc} = \tfrac{1}{2}(c^\dagger_{abc}+c^\dagger_{bca}-c^\dagger_{cab})
-\tfrac{1}{2}({\cal T}^\dagger_{abc}+{\cal T}^\dagger_{bca}-{\cal
  T}^\dagger_{cab}).
\label{eafromht}
\ee

As in our discussion of WGT, it is convenient to list the Bianchi
identities satisfied by the gravitational gauge field strengths
${{\cal R}^{\dagger ab}}_{cd}$, ${{\cal T}^{\dagger a}}_{bc}$ and
${\cal H}^\dagger_{ab}$ in eWGT. These may again be straightforwardly
derived from the Jacobi identity, but now applied to the eWGT covariant
derivative. One quickly finds the three Bianchi identities
\cite{eWGTpaper}
\begin{subequations}
\begin{eqnarray}
{\cal D}^\dagger_{[a}{{\cal R}^{\dagger de}}_{bc]}-{{\cal T}^{\dagger
    f}}_{[ab} {{\cal R}^{\dagger de}}_{c]f} & = & 0, \label{ewgtbi1} \\
{\cal D}^\dagger_{[a}{{\cal T}^{\ast d}}_{bc]}-{{\cal T}^{\dagger
    e}}_{[ab} {{\cal T}^{\dagger d}}_{c]e}-{{\cal R}^{\dagger d}}_{[abc]} - {\cal H}^\dagger_{[ab}\delta^d_{c]}
 & = & 0, \label{ewgtbi2} \\
{\cal D}^\dagger_{[a}{{\cal H}}^\dagger_{bc]} -{{\cal T}^{\dagger
    e}}_{[ab} {{\cal H}}^\dagger_{c]e} & = & 0. \label{ewgtbi3}
\end{eqnarray}
\end{subequations}
By contracting over various indices, one also obtains the
following non-trivial contracted Bianchi identities:
\begin{subequations}
\begin{eqnarray}
{\cal D}^\dagger_{a}{{\cal R}^{\dagger ae}}_{bc}-2
{\cal D}^\dagger_{[b}{{\cal R}^{\dagger e}}_{c]}
-2{{\cal T}^{\dagger
    f}}_{a[b} {{\cal R}^{\dagger ae}}_{c]f}
-{{\cal T}^{\dagger
    f}}_{bc} {{\cal R}^{\dagger e}}_{f}
 & = &   0, \label{ewgtcbi1} \\
{\cal D}^\dagger_{a}({{\cal R}^{\dagger a}}_{c}
-\tfrac{1}{2}\delta^a_c{\cal R}^\dagger)
+{{\cal T}^{\dagger
    f}}_{bc} {{\cal R}^{\dagger b}}_{f}
+\tfrac{1}{2}{{\cal T}^{\dagger
    f}}_{ab} {{\cal R}^{\dagger ab}}_{cf}
& = &  0, \label{ewgtcbi2} \\
{\cal D}^\dagger_{a}{{\cal T}^{\dagger a}}_{bc}
+2{{\cal R}}^\dagger_{[bc]}
- 2{\cal H}^\dagger_{bc}
& = & 0, \label{ewgtcbi3}
\end{eqnarray}
\end{subequations}
which are somewhat simpler than their WGT counterparts
(\ref{wgtcbi1}--\ref{wgtcbi3}) on account of the condition ${\cal
  T}^\dagger_a =0$.

\subsection{Manifestly covariant variational derivatives in eWGT}

As in WGT, we begin by considering directly the variation of the
action. In particular, by analogy with (\ref{eq:hdL}), one may
immediately write
\be
h\,\delta_0{\cal L} = 
\frac{\bar{\partial} L}{\partial\vpsi_A}\,\delta_0\vpsi_A + 
\pd{L}{({\cal D}^\dagger_a\vpsi_A)}\,\delta_0({\cal D}^\dagger_a\vpsi_A) +
\pd{L}{{\cal R}^\dagger_{abcd}}\,\delta_0{\cal R}^\dagger_{abcd} +
\pd{L}{{\cal T}^\dagger_{abc}}\,\delta_0{\cal T}^\dagger_{abc} +
\pd{L}{{\cal H}^\dagger_{ab}}\,\delta_0{\cal H}^\dagger_{ab}- {b^a}_\mu L\,\delta_0 {h_a}^\mu.
\label{eq:hdLewgt}
\ee
In eWGT, however, there is an additional subtlety compared with WGT:
although the dynamical energy-momentum tensor ${{\tau}^a}_\mu \equiv
\delta {\cal L}/\delta {h_a}^\mu$ derived from the {\it total}
Lagrangian density is covariant, this does {\it not} necessarily hold
for the corresponding quantities obtained from {\it subsets} of the
terms in $L$, even if they transform covariantly with weight $w = -4$
\cite{eWGTpaper}. This leads one to the construct an alternative
quantity for which this more general covariance property does
hold. This may be arrived at more directly from an alternative
variational principle, in which one makes a change of field variables
from the set $\vpsi_A$, $h_a^\mu$, ${A^{ab}}_\mu$ and $B_\mu$ to the
new set $\vpsi_A$, $h_a^\mu$, ${A^{\dagger ab}}_\mu$ and $B_\mu$.  It
is worth noting that one is simply making a change of field variables
here, rather than considering ${A^{\dagger ab}}_\mu$ to be an
independent field variable; in other words, one still considers
${A^{\dagger ab}}_\mu$ to be given in terms of $h_a^\mu$,
${A^{ab}}_\mu$, $B_\mu$ by its defining relationship
(\ref{adaggerdef}), rather than an independent quantity whose
relationship to the other variables would be determined from the
variational principle.
Moreover, as shown in \cite{eWGTpaper}, the eWGT covariant derivative
can be expressed wholly in terms of the fields $h_a^\mu$ (or its
inverse) and ${A^{\dagger ab}}_\mu$, and thus so too can the eWGT
field strengths. In particular, if one defines the (non-covariant)
derivative operator ${\cal D}^\natural_a\vpsi \equiv {h_a}^\mu
D^\natural_\mu \vpsi \equiv {h_a}^\mu (\partial_\mu +
\tfrac{1}{2}{A^{\dagger bc}}_\mu\Sigma_{bc})\vpsi$ and the quantities
${{\cal T}^{\natural a}}_{bc} \equiv 2{h_b}^\mu {h_c}^\nu
D^\natural_{[\mu} {b^a}_{\nu]}$, then one may easily show that ${\cal
  D}^\dagger_a\vpsi = ({\cal D}^\natural_a-\tfrac{1}{3}w{\cal
  T}^\natural_a)\vpsi$. Consequently, in the new set of field
variables, the Lagrangian $L$ in (\ref{eq:ewgtlag}) has no explicit
dependence on $B_\mu$.

Following the general procedure used for WGT, one must now determine
how the variations in (\ref{eq:hdLewgt}) depend on the variations the
new set of fields $\vpsi_A$, $h_a^\mu$ and ${A^{\dagger ab}}_\mu$
themselves. This is easily achieved using the definition of the eWGT
covariant derivative and the expressions
(\ref{erfsdef}--\ref{etfsdef}) for the field strengths.
By analogy with the approach adopted for WGT, one must also 
make use of the fact that for any coordinate vector $V^\mu$ of
weight $w=0$ (i.e.\ invariant under local scale transformations, like
the Lagrangian density ${\cal L}$), one may show that $\partial_\mu
V^\mu = h^{-1}{\cal D}^\dagger_a(h{b^a}_\mu V^\mu)$
or, equivalently, for any local Lorentz vector ${\cal V}^a$ having
Weyl weight $w = -3$ one has \cite{eWGTpaper}
\be
{\cal D}^\dagger_a{\cal V}^a = h\partial_\mu(h^{-1}
   {h_a}^\mu {\cal V}^a),
\label{eq:totaldewgt}
\ee
which is somewhat simpler than its WGT counterpart
(\ref{eq:totaldewgt}) because of the condition ${\cal T}^\dagger_a
=0$. Expressions of the
form (\ref{eq:totaldewgt}) on the RHS of (\ref{eq:hdLewgt}) therefore
contribute only surface terms to the variation of the action in
(\ref{eq:genactioninv-mancov}), but we will retain them nonetheless,
as they are required for our later discussion.

We begin by
considering together the first two terms on the RHS of
(\ref{eq:hdLewgt}), for which one obtains (after a rather lengthy
calculation)
\bea
\frac{\bar{\partial} L}{\partial\vpsi_A}\,\delta_0\vpsi_A + 
\pd{L}{({\cal D}^\dagger_a\vpsi_A)}\,\delta_0({\cal D}^\dagger_a\vpsi_A)
& & \nonumber\\
& & \hspace*{-4.5cm} = \frac{\bar{\partial} L}{\partial\vpsi_A}\,\delta_0\vpsi_A +
\pd{L}{({\cal D}^\dagger_a\vpsi_A)}\left[{\cal
    D}^\dagger_a(\delta_0\vpsi_A)+\delta_0 {h_a}^\mu D^\dagger_\mu\vpsi_A
+(\tfrac{1}{2}{h_a}^\mu\Sigma_{bc}+\tfrac{1}{3}w_A\eta_{a[c}{h_{b]}}^\mu)\vpsi_A\,\delta_0{A^{\dagger
    bc}}_\mu\right. \nonumber\\
&&\hspace{3.4cm}\left.+\tfrac{2}{3}w_A\vpsi_A({h_{[a}}^\mu{\cal
    D}^\dagger_{b]}+\tfrac{1}{2}{h_c}^\mu{{\cal T}^{\dagger c}}_{ab})\,\delta_0{b^b}_\mu\right],\nonumber\\
&& \hspace*{-4.5cm} = \left[\frac{\bar{\partial} L}{\partial\vpsi_A} 
- {\cal D}^\dagger_a \pd{L}{({\cal D}^\dagger_a\vpsi_A)}\right]\delta_0\vpsi_A
+ \left[\pd{L}{({\cal D}^\dagger_a\vpsi_A)}D^\dagger_\mu\vpsi_A + 
\tfrac{2}{3}w_A{b^c}_\mu\delta^a_{[b}{\cal D}^\dagger_{c]}\left(\pd{L}{({\cal D}^\dagger_b\vpsi_A)}\vpsi_A\right)
\right]\,\delta_0{h_a}^\mu \nonumber \\
&& \hspace{0.35cm} + \pd{L}{({\cal D}^\dagger_a\vpsi_A)}(\tfrac{1}{2}{h_a}^\mu\Sigma_{bc}+\tfrac{1}{3}w_A\eta_{a[c}{h_{b]}}^\mu)\vpsi_A\,\delta_0{A^{\dagger
    bc}}_\mu,\nonumber \\
&& \hspace{0.35cm} + \,{\cal D}^\dagger_a\left[\pd{L}{({\cal
      D}^\dagger_a\vpsi_A)}\delta_0\vpsi_A
  +\tfrac{2}{3} \pd{L}{({\cal
      D}^\dagger_{[a}\vpsi_A)} w_A \vpsi_A {b^{b]}}_\mu\,\delta_0{h_b}^\mu\right],
\label{eq:term1+2ewgt}
\eea
where both terms in square brackets in the last line are
readily shown to have Weyl weight $w=-3$, with the second one
having no analogue in the
corresponding expression (\ref{eq:term1+2}) in WGT.
Analysing the further terms containing derivatives on the RHS of
(\ref{eq:hdLewgt}) in a similar manner, one finds (again after lengthy
calculations in each case)
\bea
\pd{L}{{\cal R}^\dagger_{abcd}}\,\delta_0{\cal R}^\dagger_{abcd}
& = & 2\pd{L}{{\cal R}^\dagger_{abcd}}\left[R^\dagger_{ab\mu d}\,\delta_0 {h_{c}}^\mu
+ {h_{d}}^\mu{\cal D}^\dagger_{c}(\delta_0 A^\dagger_{ab\mu})\right], \nonumber\\
& = & 2\pd{L}{{\cal R}^\dagger_{abcd}}R^\dagger_{ab[\mu d]}\,\delta_0{h_c}^\mu
\! + \! \left( {h_e}^\mu {{\cal T}^{\dagger e}}_{cd}+
2{h_c}^\mu{\cal D}^\dagger_d 
\right)\left(\pd{L}{{\cal R}^\dagger_{abcd}}\right)\,\delta_0
A^\dagger_{ab\mu}
\!-\! 2{\cal D}^\dagger_d\left[ \pd{L}{{\cal R}^\dagger_{abcd}}{h_c}^\mu\,\delta_0 A^\dagger_{ab\mu}\right],\phantom{AAa}
\label{eq:eterm3}\\[3mm]
\pd{L}{{\cal T}^\dagger_{abc}}\,\delta_0{\cal T}^\dagger_{abc} 
& = & 2\pd{L}{{\cal T}^\dagger_{abc}}\left[T^\dagger_{a\mu\nu}{h_c}^\nu\,\delta_0{h_b}^\mu
+ {h_c}^\nu {\cal D}^\dagger_b(\delta_0 b_{a\nu}) +
{h_b}^\mu\,\delta_0 A^\dagger_{ac\mu}\right.\nonumber \\
&&\hspace{3.4cm}
-\tfrac{1}{3}\left.\eta_{ac}(\eta_{b[p}{h_{q]}}^\mu\,\delta_0{A^{\dagger
    pq}}_\mu+2{h_{[q}}^\mu{\cal D}^\dagger_{b]}(\delta_0{b^q}_\mu)
+ {h_p}^\mu{{\cal T}^{\dagger p}}_{qb}\,\delta_0{b^q}_\mu)\right],
\nonumber \\
& = & 2\pd{L}{{\cal T}^\dagger_{abc}}\!
\left[(T^\dagger_{a\mu\nu}{h_c}^\nu \delta_b^d-\tfrac{1}{2}{{\cal
      T}^{\dagger d}}_{bc}b_{a\mu})\delta_0{h_d}^\mu + {h_b}^\mu\delta_0 A^\dagger_{ac\mu}
\right]
- 2{\cal D}^\dagger_c
\left(\pd{L}{{\cal T}^\dagger_{abc}}\right)b_{a\mu}\,\delta_0
     {h_b}^\mu \nonumber \\
&& \hspace{1.8cm} -\tfrac{2}{3}\eta_{ac}
\left[({b^p}_\mu{\cal D}^\dagger_b - \delta^p_b {b^q}_\mu {\cal
    D}^\dagger_q)\left(\pd{L}{{\cal T}^\dagger_{abc}}\right)\delta_0 {h_p}^\mu
+\pd{L}{{\cal T}^\dagger_{abc}}\eta_{b[p}{h_{q]}}^\mu\,\delta_0{A^{\dagger pq}}_\mu
\right] \nonumber \\
&&\hspace{1.8cm} + 2{\cal D}^\dagger_c
\left[\left(\pd{L}{{\cal T}^\dagger_{abc}}b_{a\mu}
-\tfrac{2}{3}\eta_{pq}\pd{L}{{\cal T}^\dagger_{pq[c}} {b^{b]}}_\mu
  \right)
  \delta_0
  {h_b}^\mu\right],
\phantom{AA}
\label{eq:eterm4} \\[3mm]
\pd{L}{{\cal H}^\dagger_{ab}}\,\delta_0{\cal H}^\dagger_{ab}
& = & 2\pd{L}{{\cal H}^\dagger_{ab}}\left[H^\dagger_{\mu\nu} {h_b}^\nu\,\delta_0 {h_a}^\mu
+{h_b}^\nu{\cal D}^\dagger_a(\delta_0 B_\nu-\tfrac{1}{3}\delta_0 T_\nu)\right], \nonumber \\
& = & 2\pd{L}{{\cal H}^\dagger_{ab}} H^\dagger_{\mu\nu}{h_b}^\nu\,\delta_0
   {h_a}^\mu + \tfrac{2}{3}{b^a}_{[\mu}{\cal D}^\dagger_{c]}
\left[({{\cal T}^{\dagger c}}_{pq} + 2\delta_p^c{\cal
     D}^\dagger_q)\left(\pd{L}{{\cal
    H}^\dagger_{pq}}\right)
\right]\delta_0{h_a}^\mu
 \nonumber \\
&& \hspace{3.2cm} +\tfrac{2}{3}\eta_{c[a}{h_{b]}}^\mu(\delta_p^c{\cal
     D}^\dagger_q + \tfrac{1}{2}{{\cal T}^{\dagger c}}_{pq})\left(\pd{L}{{\cal
     H}^\dagger_{pq}}\right)\delta_0{A^{\dagger ab}}_\mu\nonumber \\
&&\hspace{3.2cm} -\tfrac{2}{3}{\cal D}^\dagger_c\left\{
\left[\pd{L}{{\cal H}^\dagger_{pq}} {{\cal T}^{\dagger\,[c}}_{pq}
  {b^{a]}}_\mu
+2\delta_p^{[c}{\cal D}^\dagger_q \left(\pd{L}{{\cal H}^\dagger_{pq}}\right) {b^{a]}}_\mu
\right]
\delta_0 {h_a}^\mu   
 \right\}.
\label{eq:eterm5}
\eea
In the above expressions it is again assumed that the appropriate
antisymmetrisations, arising from the symmetries of the field strength
tensors, are performed when the RHS are evaluated. It is also easily
shown that the quantity in brackets in each of the last terms
in (\ref{eq:eterm3}--\ref{eq:eterm5}) has Weyl weight $w=-3$, so
according to (\ref{eq:totaldewgt}) each
such term contributes a surface term to the variation of the action
(\ref{eq:genactioninv-mancov}).

Following an analogous approach to that adopted for WGT, one may
then substitute the expressions (\ref{eq:term1+2ewgt}--\ref{eq:eterm5})
into (\ref{eq:hdLewgt}), which may itself subsequently be substituted into
(\ref{eq:genactioninv-mancov}) to obtain an expression of the general
form (\ref{eq:genactioninv-mancov2}) for Noether's first theorem.
This may be written as
\be
\delta S = \int \left[\upsilon^A\,\delta_0\vpsi_A
+ {\tau^{\dagger a}}_\mu\,\delta_0{h_a}^\mu 
+ {\sigma_{ab}}^\mu\,\delta_0{A^{\dagger ab}}_\mu
+ h^{-1}{\cal D}^\dagger_p (h{\cal J}^p)
\right]\,d^4x = 0,
\ee
where the current $h{\cal J}^p$ is given by
\bea
h{\cal J}^p &=& \pd{L}{({\cal
    D}^\dagger_p\vpsi_A)}\delta_0\vpsi_A\nonumber \\
&&+2\left[\tfrac{1}{3} \pd{L}{({\cal
      D}^\dagger_{[p}\vpsi_A)} w_A \vpsi_A {b^{b]}}_\mu + \pd{L}{{\cal T}^\dagger_{abp}}b_{a\mu}
-\tfrac{2}{3}\eta_{rs}\pd{L}{{\cal T}^\dagger_{rs[p}} {b^{b]}}_\mu
-\tfrac{1}{3}\pd{L}{{\cal H}^\dagger_{rs}} {{\cal T}^{\dagger\,[p}}_{rs}
  {b^{b]}}_\mu
-\tfrac{2}{3}\delta_r^{[p}{\cal D}^\dagger_s \left(\pd{L}{{\cal H}^\dagger_{rs}}\right) {b^{b]}}_\mu
\right] \delta_0 {h_b}^\mu\nonumber \\
&&- 2\pd{L}{{\cal R}^\dagger_{abcp}}{h_c}^\mu\,\delta_0
A^\dagger_{ab\mu}
+ {b^p}_\mu \xi^\mu L,
\label{eq:ewgtjcurrent}
\eea
and we have defined the variational derivative\footnote{We denote the
variational derivative of ${\cal L}$ with respect to any one of the
fields $\chi$ in the new set of variables by $(\delta{\cal
  L}/\delta\chi)_\dagger$ to distinguish it from the variational
derivative $\delta{\cal L}/\delta\chi$ in the original set.}
$\upsilon^A \equiv (\delta{\cal L}/\delta\vpsi_A)_\dagger = \delta{\cal
  L}/\delta\vpsi_A$ with respect to the matter field $\vpsi_A$, and the
total modified dynamical energy-momentum ${\tau^{\dagger a}}_\mu
\equiv (\delta{\cal L}/\delta {h_a}^\mu)_\dagger$ and
spin-angular-momentum ${\sigma_{ab}}^\mu \equiv (\delta{\cal L}/\delta
{A^{\dagger ab}}_\mu)_\dagger = \delta{\cal L}/\delta {A^{ab}}_\mu$ of
both the matter and gravitational gauge fields.
It is also worth
noting that the (identically vanishing) dilation current $\zeta^{\dagger\mu} \equiv
(\delta{\cal L}/\delta B_\mu)_\dagger$ in the new set of
variables is related to that in the original set by $\zeta^{\dagger\mu}
= \zeta^\mu-2{h_a}^\mu{\sigma^{ab}}_b$, so that the latter is
given simply by  $\zeta^\mu=2{h_a}^\mu{\sigma^{ab}}_b$.
Manifestly covariant forms for the variational derivatives may then be read
off from the expressions
(\ref{eq:term1+2ewgt}--\ref{eq:eterm5}). Converting all Greek indices
to Roman and defining the quantities ${\tau^{\dagger a}}_b \equiv
{\tau^{\dagger a}}_\mu {h_b}^\mu$ and ${\sigma_{ab}}^c \equiv
{\sigma_{ab}}^\mu{b^c}_\mu$, one then makes the following identifications
%
%
%
%
%
\begin{subequations}
\label{eq:ewgtmcvard}
\bea
h\upsilon^A &=& \frac{\bar{\partial} L}{\partial\vpsi_A} 
- {\cal D}^\dagger_a\pd{L}{({\cal D}^\dagger_a\vpsi_A)}, \label{eq:eeomvpsi}\\[3mm]
h{\tau^{\dagger a}}_b & = & \pd{L}{({\cal D}^\dagger_a\vpsi_A)}{\cal D}^\dagger_b\vpsi_A
\!+\! 2\pd{L}{{\cal R}^\dagger_{pqra}}{\cal R}^\dagger_{pqrb}
\!+\! 2\pd{L}{{\cal H}^\dagger_{pa}}{\cal H}^\dagger_{pb}
\!+\! 2\pd{L}{{\cal T}^\dagger_{pqa}}{\cal T}^\dagger_{pqb} 
\!-\! ({{\cal T}^{\dagger a}}_{qr} + 2\delta^a_q {\cal D}^\dagger_r)
\pd{L}{{{\cal T}^{\dagger b}}_{qr}}
\!-\!\delta_a^b L
\!-\! 2{\cal D}^\dagger_c(h{\hat{\sigma}^{ca}}_{\phantom{ca}b}),\nonumber\\[-3mm]
&&
\label{eq:eeomh}\\[2mm]
h{\sigma_{ab}}^c 
& = &  \tfrac{1}{2}\pd{L}{({\cal D}^\dagger_c\vpsi_A)}\Sigma_{ab}\vpsi_A
+\left({{\cal T}^{\dagger c}}_{rs} + 2\delta_r^c{\cal D}^\dagger_s
\right)\pd{L}{{{\cal R}^{\dagger ab}}_{rs}} -2\pd{L}{{{\cal T}^{\dagger [ab]}}_{c}}
+ h\hat{\sigma}_{ab}^{\phantom{ab}c}
\label{eq:eeoma} 
\eea
\end{subequations}
where for convenience we have also defined the quantity
\be
h\hat{\sigma}_{ab}^{\phantom{ab}c} = \tfrac{1}{3}\delta^c_{[a}\eta_{b]r}\pd{L}{({\cal
    D}^\dagger_r\vpsi_A)}w_A\vpsi_A
+\tfrac{2}{3}\eta_{pr}\delta^c_{[a}\eta_{b]q}\pd{L}{{\cal
    T}^\dagger_{pqr}}
-\tfrac{1}{3}\delta^c_{[a}\eta_{b]r}\left({{\cal T}^{\dagger r}}_{pq} + 2\delta_p^r{\cal D}^\dagger_q
\right)\pd{L}{{\cal H}^\dagger_{pq}}.
\label{eq:eeomaextra}
\ee
Once again, it is assumed that the appropriate
antisymmetrisations, arising from the symmetries of the field strength
tensors, are performed when the RHS are evaluated. As mentioned above,
$\zeta^{\dagger a} \equiv 0$ since ${\cal L}$ does not explicitly
depend on $B_\mu$ in the new set of variables; the dilation current in
the original set of variables is thus given by
$\zeta^a=2{\sigma^{ab}}_b$.  As anticipated, the expressions
(\ref{eq:eeomvpsi}--\ref{eq:eeoma}) are manifestly covariant (and
hence so too are the equations of motion obtained by setting each RHS
to zero) and straightforward to evaluate, requiring one only to
differentiate the Lagrangian $L$ with respect to the matter fields,
their covariant derivatives and the field strengths.  One may easily
confirm that the above expressions lead to precisely the same
variational derivatives as those obtained by using the standard (but
much longer) approach of evaluating (\ref{eq:eleq}) for each field.

It is worth comparing the expressions
(\ref{eq:eeomvpsi}--\ref{eq:eeoma}) with their counterparts
(\ref{eq:eomvpsi}--\ref{eq:eoma}) in WGT.  One sees that the eWGT
expression for $h\upsilon_A$ is obtained simply by `replacing
asterisks with daggers' and recalling that ${\cal T}^\dagger_a \equiv
0$, but the expressions in eWGT for $h{\tau^{\dagger a}}_b$ and
$h{\sigma_{ab}}^c$ each contain an additional final term beyond those
obtained by performing the same process on their WGT counterparts
(\ref{eq:eomh}--\ref{eq:eoma}). In particular, one sees that the final
terms in (\ref{eq:eeomh}) and (\ref{eq:eeoma}) each depend on the
quantity (\ref{eq:eeomaextra}) and have no analogue in WGT. It is a
noteworthy feature of eWGT that the additional term in the expression
for $h{\tau^{\dagger a}}_b$ is given by the covariant derivative of
the additional term (with permuted indices) in the expression for
$h{\sigma_{ab}}^c$, and this has some novel consequences.  First, one
notes that for a Lagrangian $L$ that does not contain the gauge field
strength tensors, but depends only on the matter fields and their
covariant derivatives, the variational derivatives with respect to the
gauge fields do {\it not} reduce to the {\it covariant canonical
  currents} \cite{Blagojevic02,CGTpaper} of the matter fields. Indeed,
there exist additional terms proportional to the dilational generator
$\Delta = w_A I$ for the matter fields $\vpsi_A$, so that any matter
field with non-zero Weyl weight $w_A$ contributes additionally both to
the modified energy-momentum tensor and to the spin-angular-momentum
tensor, irrespective of its spin. Second, for Lagrangians that do
depend on the gauge field strengths, there are additional terms
capable of producing a dependence on the covariant derivatives of the
field strength tensors, and in each case these terms depend on the
covariant derivatives of field strength tensors for different gauge
fields than those with respect to which the variational derivative is
taken. Moreover, the final term on the RHS of (\ref{eq:eeomh})
contains {\it second} covariant derivatives of $\partial L/\partial
{\cal H}^\dagger_{ab}$.

From (\ref{edfsdef}), it appears at first sight that ${\cal
  H}^\dagger_{ab}$ is linear in second-order derivatives of
${h_a}^\mu$ and first-order derivatives of ${h_a}^\mu$ and
${A^{\dagger ab}}_\mu$ (and hence of ${A^{ab}}_\mu$ and $B_\mu$).  In
that case, if the Lagrangian contains a term proportional to ${\cal
  H}^\dagger_{ab}{\cal H}^{\dagger\,ab}$ (which has the required Weyl
weight $w=-4$ to be scale-invariant) it would follow that the final
term on the RHS of (\ref{eq:eeomh}) is linear in fourth-order
derivatives of ${h_a}^\mu$ and third-order derivatives of all three
gauge fields ${h_a}^\mu$, ${A^{ab}}_\mu$ and $B_\mu$. Similarly, the
final term in (\ref{eq:eeoma}) would be linear in third-order
derivatives of ${h_a}^\mu$. Moreover, if the Lagrangian contains a
term proportional to ${\cal R}^\dagger_{[ab]}{\cal H}^{\dagger\,ab}$,
the final term on the RHS of (\ref{eq:eeomh}) would be linear in
third-order derivatives of ${h_a}^\mu$, ${A^{ab}}_\mu$ and $B_\mu$.
These considerations would seem to indicate that eWGTs containing
either term in the Lagrangian suffer from Ostrogradsky's instability
\cite{Woodard15,Motohashi15}. As noted in \cite{eWGTpaper}, however,
this conclusion is not clear cut, since in applying such theories to
particular physical systems or in the general linearised case, one
finds that the resulting field equations always organise themselves
into combinations of coupled second-order equations in the gauge
fields \cite{eWGTpaper}. Specifically, one finds the terms containing
higher-order derivatives correspond to the derivative of already known
expressions, and so contain no new information.  Having now identified
the gauge symmetry (\ref{eq:tsg}) and obtained the general expressions
(\ref{eq:eeomh}) and (\ref{eq:eeoma}) for the variational derivatives,
one may indeed show that this always occurs in the general non-linear
case.  First, one may use the gauge transformation (\ref{eq:tsg}) to
set $T_\mu = 0$, so that ${\cal H}^\dagger_{ab}$ is merely linear in
first-order derivatives of $B_\mu$. Nonetheless, if the Lagrangian
contains a term proportional to ${\cal H}^\dagger_{ab}{\cal
  H}^{\dagger\,ab}$, the final term in (\ref{eq:eeomh}), specifically
the part that arises from the final term in (\ref{eq:eeomaextra}),
still contains third-order derivatives of $B_\mu$. This is
unproblematic, however, since this term is the covariant derivative of
an expression that is already known from the field equation
$h{\sigma_{ab}}^c = 0$. Hence, in the final field equations one
encounters field derivatives of only second-order or lower, thereby
avoiding Ostrogradsky's instability.

It is also worth pointing out that, as for WGT, we have not assumed
the equations of motion to be satisfied in deriving
(\ref{eq:eeomvpsi}--\ref{eq:eeoma}). Thus, one may calculate the
corresponding variational derivatives for {\it any subset} of terms in
$L$ that is a scalar density of weight $w=-4$. Individually, however,
such quantities do {\it not} vanish, in general. Rather, each equation
of motion requires only the vanishing of the sum of such quantities,
when derived from disjoint subsets that exhaust the total Lagrangian
$L$.

\subsection{Relationship between first- and second-order variational
  principles in eWGT}

As we did for WGT, we now demonstrate how the approach outlined above
is well suited to comparing first- and second-order variational
derivatives. We again focus on the example of the variational
derivatives obtained by setting the (eWGT) torsion to zero {\it after}
the variation is performed (first-order approach) with those obtained
by setting the torsion to zero in the action {\it before} carrying out
the variation (second-order approach). As mentioned in the
Introduction, however, in eWGT one faces an additional complication
relative to WGT, since setting the torsion to zero does not lead to an
explicit expression for the rotational gauge field in terms the other
gauge fields, but instead an implicit constraint relating all the
gauge fields.

We again begin by considering the simpler case of the first-order
approach, where one merely sets ${{\cal T}^{\dagger a}}_{bc} = 0$
(which is a properly eWGT-covariant condition) in the expressions
(\ref{eq:eeomvpsi}--\ref{eq:eeoma}). In eWGT, however, the
condition ${{\cal T}^{\dagger a}}_{bc} = 0$ results in an {\it
  implicit} constraint between the gauge fields ${h_a}^\mu$,
${A^{ab}}_\mu$ and $B_\mu$.  Once again, it proves useful in eWGT to
work in terms of the modified rotational gauge field, or rather its
`reduced' form in the case ${{\cal T}^{\dagger a}}_{bc} = 0$
\cite{eWGTpaper,CGTpaper}. From (\ref{eafromht}), this is
given by $\zero{A}^\dagger_{ab\mu} = {b^c}_\mu\czero{{\cal
    A}}^\dagger_{abc}$, where\footnote{It is important to note that
there is a fundamental difference with WGT here, since
${\zero{A}^{\dagger ab}}_{\mu}$ depends on the rotational gauge field
${A^{ab}}_\mu$ through the terms containing ${\cal T}_a$, and hence
cannot be written entirely in terms of the other gauge fields
${h_a}^\mu$ and $B_\mu$.}
\be
\zero{{\cal A}^\dagger_{abc}} =
\tfrac{1}{2}(c_{abc}+c_{bca}-c_{cab})+\eta_{ac}({\cal B}_b
-\tfrac{1}{3}{\cal T}_b)
-\eta_{bc}({\cal B}_a-\tfrac{1}{3}{\cal T}_a).
\label{eq:adaggerzero}
\ee
In an analogous manner to WGT, under a local extended Weyl
transformation, the quantities $\zero{{A^{\dagger ab}}_\mu}$ transform
in the same way as ${A^{\dagger ab}}_\mu$, and so one may construct
the `reduced' eWGT covariant derivative $\czero{{\cal
    D}}^\dagger_a\vpsi = {h_a}^\mu \,\zero{D^\ast_\mu} \vpsi =
{h_a}^\mu (\partial_\mu + \tfrac{1}{2}\zero{{A^{\dagger
      ab}}_\mu}\Sigma_{ab}+wB_\mu)\vpsi$, which transforms in the same
way as ${\cal D}^\dagger_a\vpsi$. Thus, the
corresponding quantities to (\ref{eq:eeomvpsi}--\ref{eq:eeoma}) are
obtained simply by evaluating the RHS with ${{\cal T}^{\dagger a}}_{bc}$
set to zero, which also implies ${\cal
  D}^\dagger_a \to \czero{{\cal D}}^\dagger_a$. This yields
\begin{subequations}
\label{eq:weeom0}
\bea
h\,\zero{\upsilon}^A &=& 
\left.\frac{\bar{\partial} L}{\partial\vpsi_A}\right|_0 
- \czero{{\cal D}}^\dagger_a
\left.\pd{L}{({\cal D}^\dagger_a\vpsi_A)}\right|_0, \label{eq:eeomvpsi0}\\[3mm]
h\,\zero{{\tau^{\dagger a}}_b} & = & 
\left.\pd{L}{({\cal D}^\dagger_a\vpsi_A)}\right|_0
\czero{{\cal D}}^\dagger_b\vpsi_A
+ 2\left.\pd{L}{{\cal R}^\dagger_{pqra}}\right|_0\czero{{\cal R}}^\dagger_{pqrb}
+ 2\left.\pd{L}{{\cal H}^\dagger_{pa}}\right|_0{\cal H}^\dagger_{pb}
 + 2\,\czero{{\cal D}}^\dagger_r\left.\pd{L}{{{\cal
      T}^{\dagger
       b}}_{ar}}\right|_0-\delta_a^b \left.L\right|_0
-2\,\czero{{\cal D}}^\dagger_c(h\,\zero{\hat{\sigma}}^{ca}_{\phantom{ca}b}),
\phantom{AAAa}
\label{eq:eeomh0}\\[3mm]
h\,\zero{{\sigma_{ab}}^c} 
& = &  \tfrac{1}{2}
\left.\pd{L}{({\cal D}^\dagger_c\vpsi_A)}\right|_0 \Sigma_{ab}\vpsi_A
+2\delta_r^c\,\czero{{\cal D}}^\dagger_s
\left.\pd{L}{{{\cal R}^{\dagger ab}}_{rs}}\right|_0 
-2\left.\pd{L}{{{\cal T}^{\dagger [ab]}}_{c}}\right|_0
+ h\,\zero{\hat{\sigma}}_{ab}^{\phantom{ab}c},
\label{eq:eeoma0}
\eea
\end{subequations}
where by analogy with (\ref{eq:eeomaextra}) we have defined the quantity
\be
h\,\zero{\hat{\sigma}}_{ab}^{\phantom{ab}c}
=\tfrac{1}{3}\delta^c_{[a}\eta_{b]r}
\left.\pd{L}{({\cal D}^\dagger_r\vpsi_A)}\right|_0 w_A\vpsi_A
+\tfrac{2}{3}\delta^c_{[a}\eta_{b]q}\eta_{pr}\left.\pd{L}{{\cal
    T}^\dagger_{pqr}}\right|_0
-\tfrac{2}{3}\delta^c_{[a}\eta_{b]p}\czero{{\cal D}}^\dagger_q
\left.\pd{L}{{\cal H}^\dagger_{pq}}\right|_0.
\label{eq:eeoma0extra}
\ee
Once again, it is worth noting that we have not assumed any equations
of motion to be satisfied in deriving the quantities
(\ref{eq:eeomvpsi0}-- \ref{eq:eeoma0}). Thus, one may derive
corresponding quantities for {\it any subset} of terms in $L$ that are
a scalar density with weight $w=-4$, and these quantities do not
vanish, in general.

We now consider the second-order approach, where one imposes ${\cal
  T}^{\dagger}_{abc} = 0$ at the level of the action, prior to
evaluating the variational derivatives. In this case, ${A^{\dagger
    ab}}_\mu$ is again given by (\ref{eq:adaggerzero}), in which case one
may show that the following constraint must be satisfied while
performing the variation:
\be
C_{ab\mu} \equiv A^\dagger_{ab\mu}-\tfrac{2}{3}{h_d}^\nu b_{[a|\mu}
  {A^{\dagger d}}_{|b]\nu}
-\zero{A}_{ab\mu}+\tfrac{2}{3}{h_d}^\nu b_{[a|\mu}
  \zero{{A^{d}}_{|b]\nu}} = 0,
\label{eq:aconstraint}
\ee
where $\zero{A}_{ab\mu} =
\tfrac{1}{2}{b^c}_\mu(c_{abc}+c_{bca}-c_{cab})$. It is worth noting
that $C_{ab\mu}$ depends on all the gauge fields; moreover, since
$\zero{A}_{ab\mu}$ depends both on the $h$-field and its derivatives,
the expression (\ref{eq:aconstraint}) constitutes a non-holonomic
constraint. We therefore consider the {\it augmented} total Lagrangian
density $\hat{\cal L} \equiv {\cal L} + \lambda^{ab\mu}C_{ab\mu}$,
where $\lambda^{ab\mu}$ is a field of weight $w=0$ with the same
symmetries as $C_{ab\mu}$ that acts as a Lagrange multiplier.  Thus,
up to terms that are the divergence of a quantity that vanishes on the
boundary of the integration region, the integrand in the expression
(\ref{eq:genactioninv}) for the variation of the action is given by
\be
\left(\frac{\delta\hat{\cal L}}{\delta\chi_A}\right)_\dagger \!\!\delta_0\chi_A
= \upsilon^A\,\delta_0\vpsi_A
+ {\tau^{\dagger a}}_\mu\,\delta_0 {h_a}^\mu 
+ {\sigma_{ab}}^\mu\,\delta_0 {A^{\dagger ab}}_\mu
+  \lambda^{ab\mu}\,\delta_0 C_{ab\mu} +  C_{ab\mu}\,\delta_0 \lambda^{ab\mu},
\label{eq:masterewgtaug}
\ee
From (\ref{eq:aconstraint}), one finds after some calculation that
\bea
\delta_0 C_{ab\mu} &=& \delta_0 A^\dagger_{ab\mu} - 
\tfrac{2}{3}b_{[a|\mu}{h_q}^\sigma\delta_0{A^{\dagger
      q}}_{|b]\sigma}
-{b^c}_\mu \left(
{h_{[c}}^\nu\,\czero{{\cal D}^\dagger_{b]}}\delta_0 b_{a\nu}
+{h_{[a}}^\nu\,\czero{{\cal D}^\dagger_{c]}}\delta_0 b_{b\nu}
-{h_{[b}}^\nu\,\czero{{\cal D}^\dagger_{a]}}\delta_0 b_{c\nu}\right)\nonumber\\
&&\hspace{4.25cm} +\tfrac{2}{3} {b^c}_\mu
\left(\eta_{ca}{h_{[q}}^\sigma\,\czero{{\cal
      D}^\dagger_{b]}}
-\eta_{cb}{h_{[q}}^\sigma\,\czero{{\cal
      D}^\dagger_{a]}}\right)
\delta_0 {b^q}_\sigma,
\eea
from which one may show that (\ref{eq:masterewgtaug}) becomes (up to a total
divergence)
\medskip
\bea
\smash{\left(\frac{\delta\hat{\cal L}}{\delta\chi_A}\right)_\dagger}\!\!\delta_0\chi_A
&=& \zero{\upsilon}^A\,\delta_0\vpsi_A
+ \czero{{\tilde{\tau}^{\dagger a}}_{\phantom{\dagger a}\mu}}\,\delta_0 {h_a}^\mu 
+
\left({\czero{\tilde{\sigma}}_{ab}}^\mu + {\lambda_{ab}}^\mu 
-\tfrac{2}{3}{h_a}^\mu{\lambda^c}_{bc}\right)\delta_0{A^{\dagger
    ab}}_\mu \nonumber \\
&&
+ b{b^f}_\mu\left[\left(
\eta_{fa}\delta_{[b}^e\czero{{\cal D}^\dagger_{c]}}
+\eta_{fb}\delta_{[c}^e\czero{{\cal D}^\dagger_{a]}}
-\eta_{fc}\delta_{[a}^e\czero{{\cal D}^\dagger_{b]}}\right)
(h\,\lambda^{abc})
+\tfrac{4}{3}\delta^e_{[f}\czero{{\cal D}^\dagger_{b]}}(h\,{\lambda^{ab}}_a)
\right]\delta_0{h_e}^\mu
+  C_{ab\mu}\,\delta_0\lambda^{ab\mu},\phantom{AAA}
\label{eq:masterewgt0}\\
& \equiv & v^A\,\delta_0\vpsi_A
+ {t^{\dagger a}}_\mu\,\delta_0 {h_a}^\mu 
+ {s_{ab}}^\mu\,\delta_0{A^{\dagger ab}}_\mu
+  C_{ab\mu}\,\delta_0 \lambda^{ab\mu},
\label{eq:masterewgt1}
\eea
where we have again made use of (\ref{eq:totaldewgt}) and
$\czero{{\tilde{\tau}^a}_{\phantom{a}\mu}}$ and
$\czero{\tilde{\sigma}_{ab}^{\phantom{ab}c}}$ denote quantities
analogous to (\ref{eq:eeomh0}--\ref{eq:eeoma0}), respectively, but
{\it without} the terms containing $\partial L/\partial {\cal
  T}^\dagger_{abc}|_0$. In the last line, we have also defined the
modified total dynamical energy-momentum ${t^{\dagger a}}_\mu$ and
spin-angular momentum ${s_{ab}}^\mu$ of both the matter and
gravitational gauge fields, and the matter field variational
derivatives $v^A$, in the second-order approach. 

From (\ref{eq:masterewgt1}), one sees immediately that the equation of
motion for the Lagrange multiplier field $\lambda^{ab\mu}$ is simply
$C_{ab\mu}=0$, which enforces the original constraint
(\ref{eq:aconstraint}), as required. By comparing
(\ref{eq:masterewgt0}) and (\ref{eq:masterewgt1}), and converting all
indices to Roman, one further finds that the second-order variational
derivatives are related to the first-order ones by
\bea
hv^A & = & h\,\zero{\upsilon}^A,\label{eq:eeomspsi}\\
ht^\dagger_{ab} & = & h\,\czero{{\tilde{\tau}}^\dagger_{ab}} 
+ \czero{{\cal D}}^\dagger_c
\left(h{\lambda^c}_{ab} + h{\lambda^c}_{ba} - h{\lambda_{ab}}^c\right)
-\tfrac{2}{3}\eta_{ab}\czero{{\cal D}}^\dagger_c(h{\lambda^{cd}}_d)
+\tfrac{2}{3}\,\czero{{\cal D}}^\dagger_b(h{\lambda_{ad}}^d),\label{eq:eeomsh}\\
hs_{abc} & = & h\left(\czero{\tilde{\sigma}}_{abc} 
+ \lambda_{abc} +\tfrac{2}{3}\eta_{c[a}{\lambda_{b]d}}^d
\right).
\label{eq:eeomsa}
\eea
%
To proceed further, one must eliminate the dependence of
(\ref{eq:eeomsh}--\ref{eq:eeomsa}) on the Lagrange multiplier field
$\lambda_{abc}$. This is achieved by enforcing the $A$-field equation
of motion, so that $hs_{abc} = 0$, which now merely determinines
$\lambda_{abc}$ under the constraint $C_{ab\mu} = 0$.  Using the
resulting condition $\czero{\tilde{\sigma}}_{abc} + \lambda_{abc}
+\tfrac{2}{3}\eta_{c[a}{\lambda_{b]d}}^d = 0$, one may now eliminate
the Lagrange multiplier field from (\ref{eq:eeomsh}), and one
finally obtains
\bea
hv^A & = & h\,\zero{\upsilon}^A,\label{eq:eeomspsi2}\\
ht^\dagger_{ab} & = & h\,\czero{{\tilde{\tau}}^\dagger_{ab}} 
+ \czero{{\cal D}}^\dagger_c
\left(h\,{\czero{\tilde{\sigma}}_{ab}}^c
-h\,{\czero{\tilde{\sigma}}^{c}}_{ab}
-h\,{\czero{\tilde{\sigma}}^{c}}_{ba}\right).\label{eq:eeomsh2}
\eea
As was the case for WGT, the forms of the matter variational
derivatives are identical in the first- and second-order approaches,
and the form for the modified energy-momentum tensor in ths
second-order approach is reminiscent of the Belinfante tensor.  Since,
one has not used the equations of motion for the matter fields and the
gauge field ${h_a}^\mu$ in deriving the expressions
(\ref{eq:eeomspsi2}--\ref{eq:eeomsh2}), they remain valid for any
subset of the terms in ${\cal L}$ that are a scalar density of weight
$w=-4$.  If one does consider the total Lagrangian $L$, however, then
the second-order equations of motion for the matter and gauge fields
are obtained simply by setting the expressions
(\ref{eq:eeomspsi2}--\ref{eq:eeomsh2}) to zero. In this case, provided
the terms of the form $\partial L/\partial {\cal T}^\dagger_{abc}|_0$
vanish in the first-order equations of motion obtained by setting
(\ref{eq:eeom0}--\ref{eq:eomb0}) to zero, then this implies that the
second-order equations of motion obtained by setting
(\ref{eq:eeomspsi2}-\ref{eq:eeomsh2}) to zero are also satisfied, but
the contrary does not necessarily hold.

\subsection{Manifestly covariant conservation laws in eWGT}
\label{sec:mcconsewgt}

We now derive the conservation laws for eWGT in a manner that
maintains manifest covariance throughout, by applying the general
method outlined in Section~\ref{sec:mancovvp} in a similar way to that
performed in Section~\ref{sec:mcconswgt} for WGT. Once again, we begin
by considering the general form of the conservations laws given in
(\ref{eq:n2cond1-mancov}). As in the previous section, we work in the
new set of variables $\vpsi_A$, $h_a^\mu$, ${A^{\dagger ab}}_\mu$, in
which the Lagrangian does not depend explicitly on the gauge field
$B_\mu$. In this case, under
infinitesimal local Weyl transformations consisting of GCTs, rotations
of the local Lorentz frames and dilations, parameterised by
$\xi^\mu(x)$, $\omega^{ab}(x)$ and $\rho(x)$, the form variations
(\ref{eq:ewgtfieldtrans}) are replaced by
\begin{subequations}
\label{eq:ewgtfieldtransnew}
\bea
\delta_0\vpsi & = & -\xi^\nu\partial_\nu\vpsi +
(\tfrac{1}{2}\omega^{ab}\Sigma_{ab} + w\rho)\vpsi,
\label{eq:evpsitransnew}\\
\delta_0{h_a}^\mu & = & 
-\xi^\nu\partial_\nu {h_a}^\mu + {h_a}^\nu\partial_\nu\xi^\mu
- ({\omega^b}_{a} +\rho\,\delta^b_a){h_b}^\mu,\phantom{AAA} 
\label{eq:ehtransnew}\\
\delta_0 {A^{\dagger\,ab}}_\mu & = &-\xi^\nu\partial_\nu{A^{\dagger\,ab}}_\mu
-{A^{\dagger ab}}_\nu \partial_\mu\xi^\nu
-2{{\omega^{[a}}_{c}A^{\dagger\,b]c}}_\mu 
-\partial_\mu\omega^{ab},
\label{eq:eatransnew}
\eea
\end{subequations}
By comparing these transformation laws with the generic form
(\ref{eq:localformvar}), one may read off the functions $f_{AC}$ and
$f_{AC}^\mu$ in the latter from the coefficients of $\{\lambda^C\} =
\{\lambda^1,\lambda^2,\lambda^3\} = \{\xi^\alpha, \omega^{ab},\rho\}$
and their partial derivatives, respectively. As anticipated, one
immediately finds that many of the functions $f_{AC}$ and $f_{AC}^\mu$
are not covariant quantities. One therefore again employs the
Bessel-Hagen method to obtain new form variations of the fields in
which the functions $f_{AC}^\mu$ are manifestly covariant, as
required, although many of the functions $f_{AC}$ may also be made so.
Following the general methodology outlined in
Appendix~\ref{app:BHmethod}, we consider separately the conservation
laws that result from the invariance of the eWGT action under
infinitesimal GCTs, local rotations and local dilations, respectively.

Considering first the infinitesimal GCTs characterised by
$\xi^\alpha(x)$ (which we take to correspond to $C=1$), one may make
use of the invariance of the action under the transformations
(\ref{eq:ewgtfieldtransnew}) for arbitrary functions $\omega^{ab}(x)$
and $\rho(x)$ by choosing them in a way that yields covariant forms
for the new functions $f_{A1}^\mu$ (and also $f_{A1}$ in this case) in
the resulting form variations. This is achieved by setting
$\omega^{ab} = -{A^{\dagger\,ab}}_\nu\xi^\nu$ and $\rho =
-(B_\nu-\tfrac{1}{3}T_\mu)\xi^\nu$ (where the minus signs are included
for later convenience), which yields transformation laws of a much
simpler form than in (\ref{eq:ewgtfieldtransnew}), given by
\begin{subequations}
\label{eq:ewgtfieldtransnew2}
\bea
\delta_0\vpsi & = & -\xi^\nu D^\dagger_\nu\vpsi,
\label{eq:evpsitransnew2}\\
\delta_0{h_a}^\mu & = & 
-\xi^\nu D^\dagger_\nu {h_a}^\mu + {h_a}^\nu\partial_\nu\xi^\mu,
\label{eq:ehtransnew2}\\
\delta_0 {A^{\dagger\,ab}}_\mu & = & \xi^\nu {R^{\dagger\,ab}}_{\mu\nu}, \label{eq:eatransnew2}
\eea
\end{subequations}
From these form variations, one may immediately read off the new forms
of the functions $f_{A1}$ and $f^\mu_{A1}$, all of which are now
manifestly covariant. Inserting these expressions into the general
form (\ref{eq:n2cond1-mancov}), one directly obtains the manifestly
covariant conservation law
\be
{\cal D}^\dagger_c(h{\tau^{\dagger\,c}}_\nu) 
- h({\sigma_{ab}}^\mu {R^{\dagger\,ab}}_{\mu\nu}  - 
{\tau^{\dagger\,a}}_\mu D^\dagger_\nu h_a^\mu - \upsilon^A D^\dagger_\nu\vpsi_A) = 0,
\ee
where $h\upsilon^A = (\delta L/\delta\vpsi_A)_\dagger = \delta L/\delta\vpsi_A$.
On multiplying through by ${h_d}^\nu$, one may rewrite the
conservation law wholly in term of quantities possessing only Roman
indices as
\be
{\cal D}^\dagger_c(h{\tau^{\dagger\,c}}_d) 
- h({\sigma_{ab}}^c {{\cal R}^{\dagger\,ab}}_{cd} 
- {\tau^{\dagger\,c}}_b {{\cal T}^{\dagger\,b}}_{cd}
- \upsilon^A {\cal D}^\dagger_d\vpsi_A) = 0.
\label{eq:ewgtcons1}
\ee

We next consider invariance of the action under infinitesimal local
Lorentz rotations characterised by $\omega^{ab}(x)$ (which we take to
correspond to $C=2$). In this case, the functions $f^\mu_{A2}$ in the
set of transformation laws (\ref{eq:ewgtfieldtransnew}) are
already manifestly covariant. One may thus insert the functions
$f^\mu_{A2}$ and $f_{A2}$ read off from (\ref{eq:ewgtfieldtransnew})
directly into the general form (\ref{eq:n2cond1-mancov}), without
employing the Bessel-Hagen method. On recalling that $\Gamma^\dagger_\beta
{\sigma_{pq}}^\beta = -{A^{\dagger\,r}}_{p\beta}{\sigma_{rq}}^\beta -
{A^{\dagger\,r}}_{q\beta}{\sigma_{pr}}^\beta$ (since ${\sigma_{ab}}^\mu$ has Weyl
weight $w=0$) one finds that the final set of terms on the LHS of
(\ref{eq:n2cond1-mancov}) vanish when $\gamma^A$ corresponds to $
h{\sigma_{ab}}^\mu$, and one immediately obtains
the manifestly covariant conservation law
\be 
{\cal D}^\dagger_c(h{\sigma_{ab}}^c) + h\tau^\dagger_{[ab]}
+ \tfrac{1}{2}h\upsilon^A\Sigma_{ab}\vpsi_A = 0.
\label{eq:ewgtcons2}
\ee

Finally, we consider invariance of the action under infinitesimal
local dilations characterised by $\rho(x)$ (which we take to
correspond to $C=3$). Once again, the relevant functions $f^\mu_{A3}$
in the set of transformation laws (\ref{eq:ewgtfieldtransnew}) are
already manifestly covariant. One may thus insert $f^\mu_{A3}$ and
$f_{A3}$ read off from (\ref{eq:ewgtfieldtransnew}) directly into the
general form (\ref{eq:n2cond1-mancov}), which immediately yields the
manifestly covariant algebraic conservation law
\be 
h{\tau^{\dagger\,c}}_c - h\upsilon^A w_A\vpsi_A = 0.
\label{eq:ewgtcons3}
\ee

It is straightforward to verify that the manifestly covariant
conservations WGT laws (\ref{eq:ewgtcons1}--\ref{eq:ewgtcons3}) have the
correct forms \cite{eWGTpaper,CGTpaper} and match those derived
(albeit at considerably greater length) using the standard form of
Noether's second theorem (\ref{eq:n2cond1}).

Before moving on to consider the further condition (\ref{eq:n2cond2}) arising
from Noether's second theorem, in the context of eWGT, we note that the
conservation law (\ref{eq:ewgtcons2}) may be used to simplify the
expression (\ref{eq:eeomsh2}) for the second-order variational
derivative with respect to ${h_a}^\mu$ in terms of first-order
variational derivatives. Imposing the condition ${\cal T}^\dagger_{abc} =
0$, the conservation law (\ref{eq:ewgtcons2}) becomes
\be
\czero{{\cal D}}^\dagger_c(h\,\czero{\tilde{\sigma}_{ab}}^{\phantom{ab}c})
+ h\,\czero{\tilde{\tau}}^\dagger_{[ab]}
+\tfrac{1}{2}h\,\czero{\tilde{\upsilon}}^A
\Sigma_{ab}\vpsi_A  =  0.
\ee
If one assumes the {\it matter} equations of motion
$\czero{\tilde{\upsilon}}^A=0$ are satisfied (or, equivalently, that
the Lagrangian $L$ does not depend on matter fields), the expression
(\ref{eq:eeomsh2}) can thus be written in the simpler and 
manifestly symmetric form
\be
ht^\dagger_{ab}  \eomm  h\,\czero{{\tilde{\tau}}^\dagger_{(ab)}} -2\,\czero{{\cal
    D}}^\dagger_c(h\,\czero{\tilde{\sigma}}^c_{\phantom{c}(ab)}).
\ee

\subsection{Relationship between currents in Noether's second theorem
  in eWGT}

We conclude this section by considering the relationship in WGT
between the two currents that appear in Noether's second theorem
(\ref{eq:n2cond2}). As discussed in Section~\ref{subsec:wgtnt2}, this
equation may be re-written as ${\cal D}^\dagger_a
[h({\cal J}^a -{\cal S}^a)] = 0$, where $h{\cal J}^a$ for eWGT is given
by (\ref{eq:ewgtjcurrent}) and the expression for $h{\cal S}^a$ may be
obtained from the general form (\ref{eq:generalscurrent}), which on
using the eWGT field variations (\ref{eq:ewgtfieldtransnew})
yields
\be
h{\cal S}^p = h\left[-\xi^\mu({\tau^{\dagger\,p}}_\mu - {\sigma_{ab}}^p
  {A^{\dagger\,ab}}_\mu) + \omega^{ab}{\sigma_{ab}}^p \right]. 
\label{eq:ewgtscurrent}
\ee
As was the case for WGT, this expression does not depend on the
variational derivatives $\upsilon^A \equiv \delta{\cal
  L}/\delta\psi_A$ with respect to the matter fields since, as
expected, the functions $f^\mu_{AC}$ vanish in this case, as can be
read off from the form variations (\ref{eq:ewgtfieldtransnew}) of the
new set of fields. Thus, in order for $h{\cal S}^p$ to vanish, it is
sufficient that just the equations of motion of the gauge fields are
satisfied. Moreover, in eWGT, the current (\ref{eq:ewgtscurrent}) also
does not depend on the dilation $\rho(x)$.

If one substitutes the form variations (\ref{eq:ewgtfieldtransnew}) of
the new set of fields into the expression (\ref{eq:ewgtjcurrent}) for
$h{\cal J}^p$, one finds after a long calculation of a similar nature
to that required in WGT, which makes careful use of
the definition (\ref{eq:dcommewgt}) of the field strength tensors, the
contracted Bianchi identity (\ref{ewgtcbi3}) and the manifestly
covariant expressions (\ref{eq:eeomh}--\ref{eq:eeoma}) for the
variational derivatives with respect to the gravitational gauge
fields, that
\be
{\cal D}^\dagger_p
(h{\cal J}^p) = 
{\cal D}^\dagger_p
\left[-\xi^\mu h({\tau^p}_q{b^q}_\mu - {\sigma_{ab}}^p
  {A^{ab}}_\mu) + \omega^{ab}h{\sigma_{ab}}^p \right]
= {\cal D}^\dagger_p (h{\cal S}^p),
\ee
thereby verifying explicitly the relationship between the two currents
that is implied by Noether's second theorem (\ref{eq:n2cond2}), as was
the case in WGT. Thus, as expected for an action that is invariant
under a set of local symmetries, this relationship contains no further
information, but nonetheless provides a useful check of the derivation
of the expressions (\ref{eq:eeomh}--\ref{eq:eeoma}). Indeed, in a
similar way to WGT, the
requirement ${\cal D}^\dagger_a [h({\cal J}^a -{\cal
    S}^a)] = 0$ from Noether's second theorem can thus be used as an
alternative (albeit rather longer) means of deriving the expressions
(\ref{eq:eomh}--\ref{eq:eomb}) for the variational derivatives with
respect to the gravitational gauge fields.

\section{Conclusions}
\label{sec:conc}

We have presented a variational principle that maintains manifest
covariance throughout when applied to the actions of gauge theories of
gravity. In particular, it directly yields field equations and
conservation laws that are manifestly covariant under the symmetries
to which the action is invariant. This is achieved by deriving
explicit manifestly covariant forms for the Euler--Lagrange variational
derivatives and Noether's theorems for a generic action of the form
typically assumed in gauge theories of gravity.

The manifestly covariant form of Noether's first theorem and the
expressions for the variational derivatives derived therefrom not only
provide a significant calculational saving relative to the traditional
method of evaluation, but also yield useful insights into their
general forms. In particular, these expressions enable one easily to
establish the relationship between the forms of variational
derivatives, and hence the field equations, obtained by applying
first- and second-order variational principles, respectively.  An
interesting case is provided by comparing the variational derivatives
obtained by setting the torsion to zero after the variation is
performed (first-order approach) with those obtained by setting the
torsion to zero in the action before carrying out the variation
(second-order approach).

The re-expression of Noether's second theorem in terms of manifestly
covariant quantities provides further utility and insights. In
particular, one may use it to derive the conservation laws obeyed by
the matter and gravitational gauge fields in a manifestly covariant
manner. This also relies on being able to express the form variations
of these fields such that at least the coefficient functions of the
derivatives of the parameters of the symmetry transformations are
manifestly covariant. This may be achieved by generalising the
approach introduced by Bessel-Hagen for electromagnetism, which is
discussed in Appendix~\ref{app:BHmethod}.  The re-expression of
Noether's second theorem further allows one straightforwardly to
verify the relationship between the two currents on which it
depends. Indeed, one may use Noether's second theorem as an
alternative (albeit somewhat longer) means of deriving manifestly
covariant forms for the variational derivatives.

The manifestly covariant variational principle is illustrated by
application to the scale-invariant Weyl gauge theory (WGT) and its
recently proposed `extended' version (eWGT), but can be
straightforwardly applied to other gravitational gauge theories with
smaller or larger symmetry groups. For WGT and eWGT, the fields in the
theory consist of a translational gauge field ${h_a}^\mu$ (with
inverse ${b^a}_\mu$), a rotational gauge field ${A^{ab}}_\mu$ and a
dilational gauge field $B_\mu$, together with some set of matter
fields $\vpsi_A$, which may include a scalar compensator field.  In
eWGT, however, it is more natural to work in terms of the alternative
set of variables $\vpsi_A$, $h_a^\mu$, ${A^{\dagger ab}}_\mu$ and
$B_\mu$, where the modified rotational gauge field ${A^{\dagger
    ab}}_\mu \equiv {A^{ab}}_\mu + 2{b^{[a}}_\mu{\cal B}^{b]}$ and
${\cal B}_a = {h_a}^\mu B_\mu$. Moreover, eWGT may be shown to be
invariant under the simultaneous `torsion-scale' gauge transformations
${A^{ab}}_\mu \to {A^{ab}}_\mu + 2{b^{[a}}_\mu{\cal Y}^{b]}$ and
$B_\mu \to B_\mu - Y_\mu$, where ${\cal Y}_a = {h_a}^\mu Y_\mu$ and
$Y_\mu$ is an arbitrary vector field; this may be used to set either
$B_\mu$ or $T_\mu$ to zero, which can considerably simplify subsequent
calculations.  The scale-invariant actions for WGT and eWGT are
further assumed to depend only on the matter fields, their covariant
derivatives and the field strength tensors of the gravitational gauge
fields.  In this case, the eWGT action in the alternative set of
variables does not depend explicitly on $B_\mu$, hence reducing by one
the number of independent variational derivatives.  As might be
expected from the above considerations, one finds a number of
similarities between WGT and eWGT, and also some important and novel
differences.

Considering first the manifestly covariant expressions for the
variational derivatives in WGT, one finds that these reduce to the
corresponding covariant canonical currents of the matter fields if the
Lagrangian does not depend on the gravitational gauge field
strengths. For Lagrangians that do depend on the gauge field
strengths, one finds that the only terms that contain the covariant
derivative of a field strength tensor depend on the field strength
tensor of the gauge field with respect to which the variational
derivative is taken. By contrast, in eWGT one finds that the
variational derivatives with respect to the translational and
modified rotational gauge fields contain additional terms beyond those
obtained by `replacing asterisks with daggers' in their WGT
counterparts. Moreover, the additional terms in the translational
variational derivative are given by the covariant derivative of the
additional terms (with permuted indices) in the expression for the
rotational variational derivative; this has some novel consequences.
First, for a Lagrangian that depends only on the matter fields and
their covariant derivatives, the variational derivatives with respect
to the gauge fields do not reduce to the covariant canonical currents
of the matter fields, but comtain additional terms proportional to the
dilational generator $\Delta = w_A I$ for the matter fields
$\vpsi_A$. Thus, any matter field with non-zero Weyl weight $w_A$
contributes additionally both to the modified energy-momentum tensor
and to the spin-angular-momentum tensor, irrespective of its
spin. Second, for Lagrangians $L$ that depend on the gauge field
strengths, there are additional terms capable of producing a
dependence on the covariant derivatives of the field strength tensors,
and in each case these terms depend on the covariant derivatives of
field strength tensors for different gauge fields than those with
respect to which the variational derivative is taken. Moreover, there
exist terms containing covariant derivatives of $\partial L/\partial
{\cal H}^\dagger_{ab}$. By using the `torsion-scale' gauge symmetry
and the manifestly covariant forms of the variational derivatives,
however, one may show that the final eWGT field equations contain field
derivatives of only second-order or lower, thereby avoiding
Ostrogradsky's instability.

On comparing the variational derivatives obtained by setting the
torsion to zero after the variation is performed (first-order
approach) with those obtained by setting the torsion to zero in the
action before carrying out the variation (second-order approach), one
finds important differences between WGT and eWGT. In both cases, the
rotational gauge field is no longer an independent field, but in WGT
it may be written explicitly in terms of the other gauge fields, whereas
in eWGT there exists an implicit constraint relating all the gauge
fields. In both cases, however, one may arrive at simple expressions
for the variational derivatives in the second-order approach in terms
of those from the first-order approach. In particular, the
translational variational derivative in the second-order approach for
WGT and eWGT is the gauge theory equivalent of the Belinfante tensor.
Moreover, in WGT the second-order dilational variational derivative
may be considered to define an associated Belinfante dilation current,
which is clearly related to the `field virial' that is relevant to the
invariance of an action under special conformal transformations.

Turning to the re-expression of Noether's second theorem, the resulting
derivations of manifestly covariant forms of the conservation laws
satisfied by the fields in WGT and eWGT, yield similar forms in both
cases for the laws corresponding to invariance under local
translations and rotations, respectively. For invariance under local
dilations, however, one finds the resulting conservation law is
differential in WGT, but algebraic in eWGT. In both WGT and eWGT, one
may also use the re-expression of Noether's second theorem to verify the
relationship between the two currents on which it depends, although in
both cases this verification requires a calculation of considerable
length. Alternatively, in each case, one may use Noether's second
theorem as an alternative (albeit considerably longer) means of
deriving manifestly covariant forms for the variational derivatives.

Whilst this paper has focussed heavily on the {\it Lagrangian}
prescription of field theory, and the associated field equations and
conservation laws, we note that the techniques developed here may
impart even stronger benefits in the {\it Hamiltonian}
formulation. Hamiltonian gauge field theory is characterised by the
presence of field-valued {\it constraints}, which encode not only the
gauge symmetries but also the whole nonlinear dynamics, as elucidated
by the consistency algorithm of Dirac and
Bergmann~\cite{Anderson1951,Bergmann1955,Dirac1958}.  The fundamental
currency of the consistency algorithm is the Poisson
bracket\footnote{More sophisticated {\it Dirac}
brackets~\cite{Castellani1982} also arise: these are equally relevant
to our discussion.}, which is a bilinear in functional variations with
respect to dynamical fields. In the context of gravitational gauge
fields, the Hamiltonian formulation is typically realised using the
so-called 3+1 or Arnowitt--Deser--Misner (ADM) technique, whereby
manifest diffeomorphism covariance is preserved despite the imposition
of a spacelike foliation.  Accordingly, the ADM Poisson bracket
presents a clear opportunity for manifestly covariant variational
methods, such as those expressed in equations (\ref{eq:mcvarderivs})
and (\ref{eq:ewgtmcvard}). The Hamiltonian demand is, if anything,
more pronounced than the Lagrangian demand. In the latter case, a
countably small collection of field equations (not including indices)
must be obtained (e.g.  {\it one} set of Einstein equations).  In the
former case and for a gravitational gauge theory, all Poisson brackets
between all constraints must be evaluated in order to classify the
gauge symmetries: this can in practice correspond to tens or hundreds
of
brackets~\cite{Yo1999,Yo2002,Blagojevic2018,Barker2021,Barker2023a}.
Separately, the variations of a constraint can be more challenging
than those of an action because: (i) the constraints are typically indexed
and always (quasi-) local, necessitating the use of smearing
functions; (ii) they may contain more terms in ADM form than the
original Lagrangian; and crucially (iii) they are of
{\it unlimited}\footnote{This is due to cumulative derivatives
arising in the course of the Dirac algorithm.} order in spatial
gradients~\cite{Barker2023b} even when the Lagrangian is second order
as assumed in~(\ref{eq:genmatteraction}).  The extension of the techniques
discussed here to the higher-order, ADM variational derivative, is
left to future work.

\begin{acknowledgments}
WEVB is supported by a Research Fellowship at Girton College, Cambridge.
\end{acknowledgments}
\vspace*{-0.2cm}
\appendix

\bigskip
\section{Bessel-Hagen method for electromagnetism}
\label{app:BHmethod}

For classical electromagnetism (EM) in Minkowski spacetime ${\cal M}$
labelled using Cartesian inertial coordinates $x^\mu$, the action is
given by $S = \int {\cal L} \,d^4x$, where the Lagrangian density
${\cal L} = -\tfrac{1}{4}F_{\mu\nu}F^{\mu\nu}$ and the (Faraday) field
strength tensor $F_{\mu\nu} = \partial_\mu A_\nu - \partial_\nu
A_\mu$, in which $A_\mu$ is the electromagnetic 4-potential (which is not
to be confused with the rotational gravitational gauge field
${A^{ab}}_\mu$ appearing throughout the main text of the paper). As is
well known, the most general infinitesimal global
coordinate transformations under which the 
EM action is invariant are the 
conformal transformations\footnote{The action is also
  invariant under finite global conformal coordinate transformations
  \cite{Cunningham10,Bateman10a,Bateman10b}; these include conformal
  inversions $x^{\prime \mu} = x^\mu/x^2$ for $x^2 \neq 0$, which are
  not connected to the identity and so are not considered here.};
these have the form $x^{\prime \mu} = x^\mu + \xi^\mu(x)$, where
\be
\xi^\mu(x) = a^\mu + {\omega^\mu}_\nu x^\nu + \rho x^\mu +
c^\mu x^2 -2 c\cdot x\, x^\mu,
\label{eq:gct}
\ee
in which the 15 infinitesimal parameters $a^\mu$, $\omega^{\mu\nu} =
-\omega^{\nu\mu}$, $\rho$ and $c^\mu$ are constants, and we use the
shorthand notation $x^2 \equiv \eta_{\mu\nu} x^\mu x^\nu$ and $c\cdot
x \equiv \eta_{\mu\nu}c^\mu x^\nu$. If the four parameters $c^\mu$
defining the so-called special conformal transformation (SCT) vanish,
then (\ref{eq:gct}) reduces to an infinitesimal global Weyl
transformation. Moreover, if the parameter $\rho$ defining the
dilation (or scale transformation) also vanishes, then (\ref{eq:gct})
further reduces to an infinitesimal global Poincar\'e transformation,
consisting of a restricted Lorentz rotation defined by the six
parameters $\omega^{\mu\nu}$ and a spacetime translation defined by
the four parameters $a^\mu$.

Under the action of any infinitesimal coordinate transformation
$x^{\prime \mu} = x^\mu + \xi^\mu(x)$, the 4-potential has the form
variation
\be
\delta^{(\xi)}_0\! A_\mu = \delta^{(\xi)}\! A_\mu -\xi^\nu\partial_\nu
A_\mu = - A_\nu\partial_\mu\xi^\nu -\xi^\nu\partial_\nu A_\mu,
\label{eq:formvargct}
\ee
where we have explicitly denoted the form and total variations as
being induced by the infinitesimal coordinate transformation.  Thus,
the corresponding Noether current (\ref{eq:noethercurrent}) has the
form
\be
J^\mu 
= \momf{\mu}{A_\sigma}\delta^{(\xi)}_0\! A_\sigma + \xi^\mu{\cal L}
= F^{\mu\sigma}(A_\nu\partial_\sigma\xi^\nu +
\xi^\nu\partial_\nu A_\sigma) -\tfrac{1}{4}\xi^\mu F^{\rho\sigma}F_{\rho\sigma}.
\label{eq:emj1}
\ee
Using the expression (\ref{eq:gct}) for an infinitesimal global
conformal coordinate transformation, one finds that (\ref{eq:emj1}) 
may be written as
\be
J^\mu = -
a^\alpha {t^\mu}_\alpha + \tfrac{1}{2}\omega^{\alpha\beta}{M^\mu}_{\alpha\beta}  + \rho {D}^\mu + c^\alpha
        {{K}^\mu}_\alpha,
\label{eq:noetherj}
\ee
where the coefficients of the parameters of the conformal
transformation are defined by
\begin{subequations}
\label{eq:confcurrents}
\bea
{{t}^\mu}_\alpha &\equiv& \momf{\mu}{A_\sigma}\partial_\alpha A_\sigma-\delta^\mu_\alpha
{\cal L} = -F^{\mu\sigma}\partial_\alpha A_\sigma + \tfrac{1}{4}
\delta_\alpha^\mu F^{\rho\sigma}F_{\rho\sigma}, \label{eq:confcurrentemt}\\
{{M}^\mu}_{\alpha\beta} 
& \equiv & x_\alpha {t^\mu}_\beta - x_\beta
{t^\mu}_\alpha + {s^\mu}_{\alpha\beta}, \\
{D}^\mu & \equiv & -x^\alpha {t^\mu}_\alpha + j^\mu, \\
{{K}^\mu}_\alpha & \equiv & 
(2x_\alpha x^\beta\!-\!\delta^\beta_\alpha x^2){t^\mu}_\beta
 +  2x^\beta({s^\mu}_{\alpha\beta}\!-\!\eta_{\alpha\beta}j^\mu),
\label{eq:confcurrentsd}
\eea
\end{subequations}
which are the canonical energy-momentum, angular momentum,
dilation current and special conformal current, respectively, of the
4-potential $A_\mu$. We have also defined the quantities
\begin{subequations}
\label{eq:srcurrentsdef}
\bea
{s^\mu}_{\alpha\beta} &\equiv& 
\momf{\mu}{A_\sigma}
{(\Sigma_{\alpha\beta})_\sigma}^\rho A_\rho = -2{F^\mu}_{[\alpha}A_{\beta]},  \\
j^\mu  &\equiv&  \momf{\mu}{A_\sigma}
w A_\sigma = F^{\mu\sigma}A_\sigma,
\eea
\end{subequations}
which are the canonical spin angular momentum and intrinsic dilation
current of the 4-potential; here
${(\Sigma_{\alpha\beta})_\sigma}^\rho =
2\eta_{\sigma[\alpha}\delta_{\beta]}^\rho$ are the generators of the
vector representation of the Lorentz group and $w=-1$ is the Weyl
weight of $A_\mu$.

If the field equations $\delta {\cal L}/\delta A_\nu= \partial_\mu
F^{\mu\nu} = 0$ are satisfied, then invariance of the action implies
the conservation law $\partial_\mu J^\mu \eom 0$. Since the parameters of
the global conformal coordinate transformation in (\ref{eq:noetherj})
are constants, one thus obtains separate conservation laws  given by
\begin{subequations}
\label{eq:gctcons}
\bea
\partial_\mu {t^\mu}_\alpha & \eom & 0, \label{eq:srtransinv}\\
\partial_\mu{s^\mu}_{\alpha\beta} + 2t_{[\alpha\beta]} & \eom & 
0,\label{eq:srrotinv}\\
\partial_\mu j^\mu - {t^\mu}_\mu & \eom & 0,\label{eq:srscaleinv}\\
{s^\mu}_{\alpha\mu}-j_\alpha&\eom &0,
\label{eq:gctconsd}
\eea
\end{subequations}
which hold up to a total divergence of any quantity that vanishes on
the boundary of the integration region of the action.  It is worth
noting that the first condition has been used to derive the second and
third conditions, and the first three conditions have all been used to
derive the final condition. The conservation laws (\ref{eq:gctcons})
may be easily verified directly using the expressions
(\ref{eq:confcurrentemt}) and (\ref{eq:srcurrentsdef}) for
${t^\mu}_\alpha$, ${s^\mu}_{\alpha\beta}$ and $j^\mu$, respectively,
and the EM equations of motion. It is worth noting that the
conservation law (\ref{eq:srcurrentsdef}), which results from
invariance of the action under special conformal transformations,
requires the `field virial' to vanish \cite{Coleman71}.

In addition to being invariant under infinitesimal global conformal
coordinate transformations of the form (\ref{eq:gct}), however, the EM
action is also well known to be invariant under the gauge
transformation $A_\mu \to A'_\mu = A_\mu + \partial_\mu \alpha$, where
$\alpha(x)$ is an arbitrary function of spacetime position. Since our
considerations thus far have not taken this into account, it is
perhaps unsurprising that the canonical quantities ${t^\mu}_\alpha$,
${s^\mu}_{\alpha\beta}$ and $j^\mu$ are not invariant under the gauge
transformation, as is easily demonstrated. Moreover, it is immediately
apparent that the overall Noether current $J^\mu$ in (\ref{eq:emj1})
is also not gauge invariant. All these problems originate from the
form variation $\delta_0^{(\xi)}\! A_\sigma$ in (\ref{eq:formvargct})
itself not being gauge invariant. The lack of gauge invariance of the
canonical expressions is a severe shortcoming, which means that these
quantities must be unphysical.  The situation is usually remedied, at
least for the energy-momentum tensor in electromagnetism, by using the
Belinfante method \cite{Belinfante40} of adding ad-hoc terms, which do
not follow from Noether's theorem, to the canonical energy-momentum in
order to construct a `modified' energy-momentum tensor, which is gauge
invariant (and symmetric) and can be further `improved' to be
traceless also \cite{Callan70}.  One should note, however, that these
methods are not guaranteed to yield a gauge-invariant energy-momentum
tensor for general gauge field theories when matter fields are coupled
to a gauge field \cite{Grensing21}, although this deficiency is
addressed in \cite{Blaschke16}.

An alternative approach, which makes direct use of the gauge
invariance of the EM action and Noether's theorem, was first proposed
in 1921 by Bessel-Hagen (who acknowledges Noether for suggesting the
idea) \cite{Bessel-Hagen21}. This work is not widely known, however,
and similar approaches have since been proposed by other authors
\cite{Eriksen88,Munoz96,Burgess02,Montesinos06}, although
Bessel-Hagen's original method arguably remains the most
straightforward and intuitive \cite{Baker21}. The key to the method is
to recognise that the the form variations $\delta_0\chi_A$ of the fields
appearing in the general expression (\ref{eq:noethercurrent}) for the
Noether current $J^\mu$ may correspond to {\it any} transformation
that leaves the action invariant. Indeed, it is advantageous to
consider the {\it most general} such transformation. Applying this
notion to EM, one should thus replace the form variation
(\ref{eq:formvargct}) induced solely by the infinitesimal global
conformal coordinate transformation by the general form
\be
\delta_0 A_\mu = \delta^{(\xi)}\! A_\mu + \partial_\mu \alpha -\xi^\nu\partial_\nu
A_\mu = - A_\nu\partial_\mu\xi^\nu + \partial_\mu \alpha -\xi^\nu\partial_\nu A_\mu,
\label{eq:genformvar}
\ee
which also includes the contribution induced by the EM gauge
transformation. Since the form variation (\ref{eq:genformvar}) leaves
the EM action invariant for $\xi^\mu(x)$ given by (\ref{eq:gct}) and
for arbitrary $\alpha(x)$, one may choose the latter to be as
convenient as possible. Given that our goal is to arrive at a
gauge-invariant form for the Noether current $J^\mu$, one should
therefore choose $\alpha(x)$ such that the form variation
(\ref{eq:genformvar}) is itself gauge-invariant; this is the central
idea underlying the Bessel-Hagen method.

One may easily obtain a gauge-invariant form variation by setting
$\alpha = A_\nu\xi^\nu$, which immediately yields $\delta_0 A_\mu =
\xi^\nu F_{\mu\nu}$. Consequently, the Noether current (\ref{eq:emj1})
is replaced by the new form
\be
J^\mu 
= \momf{\mu}{A_\sigma}\delta_0 A_\sigma + \xi^\mu{\cal L}
= \xi^\nu(F^{\mu\sigma}F_{\nu\sigma} -\tfrac{1}{4}\delta_\nu^\mu
F^{\rho\sigma}F_{\rho\sigma}) = -\xi^\nu {\tau^\mu}_\nu,
\label{eq:emj2}
\ee
where in the final equality we have identified the standard physical
energy-momentum tensor ${\tau^\mu}_\nu = -(F^{\mu\sigma}F_{\nu\sigma}
-\tfrac{1}{4}\delta_\nu^\mu F^{\rho\sigma}F_{\rho\sigma})$ of the EM
field, which is immediately seen to be gauge invariant, symmetric and
traceless. Substituting the form (\ref{eq:gct}) for $\xi^\mu$ into
(\ref{eq:emj2}), one finds that the expression (\ref{eq:noetherj}) for the
Noether current is replaced by the much simpler form
\be
J^\mu = -
a^\alpha {\tau^\mu}_\alpha + \tfrac{1}{2}\omega^{\alpha\beta}
(x_\alpha {\tau^\mu}_\beta - x_\beta {\tau^\mu}_\alpha)
- \rho x^\alpha {\tau^\mu}_\alpha + c^\alpha (2x_\alpha
x^\beta-\delta_\alpha^\beta x^2){\tau^\mu}_\beta,
\label{eq:noetherjbh}
\ee
from which one can further identify new forms for the angular momentum,
dilation current and special conformal current of the EM field, all of which
are gauge invariant. If one again assumes the EM field equations
to hold and uses the fact that the
parameters of the global conformal coordinate transformation are
constants, one obtains separate conservation laws that replace those
in (\ref{eq:gctcons}) and are given by the succinct forms
\begin{subequations}
\label{eq:gctconsbh}
\bea
\partial_\mu {\tau^\mu}_\alpha & \eom & 0, \label{eq:srtransinvbh}\\
\tau_{[\alpha\beta]} & \eom &  0,\label{eq:srrotinvbh}\\
{\tau^\mu}_\mu & \eom & 0,\label{eq:srscaleinvbh}
\eea
\end{subequations}
where, in this case, the conservation law derived from the coefficient
of the SCT parameters $c^\mu$ is satisfied automatically given the
other three conservation laws above, all of which may be easily
verified directly.

Finally, one should also determine the further conservation
law that results solely from invariance of the action under EM
gauge transformations. This is easily achieved by setting $\xi^\mu =
0$, which is equivalent to all of the constant parameters in (\ref{eq:gct})
vanishing. In this case, (\ref{eq:genformvar}) becomes simply
$\delta_0 A_\mu = \partial_\mu\alpha$ and the Noether current is
immediately given by
\be
J^\mu = \momf{\mu}{A_\sigma}\delta_0 A_\sigma =
-F^{\mu\sigma}\partial_\sigma\alpha.
\ee
Assuming the EM field equations to hold, the resulting conservation
law $\partial_\mu J^\mu \eom 0$ may be written as
$F^{\mu\sigma}\partial_\mu\partial_\sigma\alpha \eom 0$, which is
satisfied identically because of the antisymmetry of
$F^{\mu\sigma}$.


\end{document}